\documentclass{emulateapj}
\usepackage{amsmath}
\usepackage{apjfonts}
\usepackage{natbib,graphicx}
\usepackage{longtable}
\newcommand{\cp}{\citep}
\newcommand{\ct}{\citet}

\begin{document}

\title{Understanding the Mass-Radius Relation for Sub-Neptunes: Radius as a Proxy for Composition}

\author{Eric D. Lopez}
\author{Jonathan J. Fortney}%\footnote{Alfred P. Sloan Research Fellow}}
\affil{Department of Astronomy and Astrophysics, University of California, Santa Cruz, CA 95064} 

\begin{abstract}
Transiting planet surveys like {\it Kepler} have provided a wealth of information on the distribution of planetary radii, particularly for the new populations of super-Earth and sub-Neptune sized planets. In order to aid in the physical interpretation of these radii, we compute model radii for low-mass rocky planets with hydrogen-helium envelopes. We provide model radii for planets 1-20 $M_{\mathrm{\oplus}}$, with envelope fractions from 0.01-20\%, levels of irradiation 0.1-1000$\times$ Earth's, and ages from 100 Myr to 10 Gyr. In addition we provide simple analytic fits that summarize how radius depends on each of these parameters. Most importantly, we show that at fixed composition, radii show little dependence on mass for planets with more than $\sim$1\% of their mass in their envelope. Consequently, planetary radius is to first order a proxy for planetary composition for Neptune and sub-Neptune sized planets. We recast the observed mass-radius relationship as a mass-{\it composition} relationship and discuss it in light of traditional core accretion theory. We discuss the transition from rocky super-Earths to sub-Neptune planets with large volatile envelopes. We suggest $1.75$ $R_{\mathrm{\oplus}}$ as a physically motivated dividing line between these two populations of planets. Finally, we discuss these results in light of the observed radius occurrence distribution found by {\it Kepler}.

\end{abstract}

\subjectheadings{planetary systems; planets and satellites: composition, formation, interiors, physical evolution}

\section{Introduction}
\label{intsec}

NASA's {\it Kepler} mission has been an enormous success, discovering over 3500 planet candidates to date \cp{Borucki2011,Batalha2012}. Among the mission's many firsts and accomplishments, however, one of the most revolutionary is that for the first time we have a robust determination of the relative abundance of different sizes of planets stretching from Earth-sized all the way up to the largest hot Jupiters \cp{Howard2011a,Fressin2013,Petigura2013}.

In particular, {\it Kepler} has discovered an abundant new population of $\sim$3 $R_{\mathrm{\oplus}}$ planets \cp{Fressin2013,Petigura2013}. Although smaller than Neptune, these planets are large enough that they must have substantial hydrogen and helium (hereafter H/He) envelopes to explain their radii. Such planets are unlike anything found in our own Solar System and fundamental questions about their structure and formation are still not understood. Are these Neptune-like planets that form beyond the snow-line and contain large amounts of volatile ices \cp{Rogers2011}, or are these scaled up terrestrial worlds with H/He envelopes that formed close to their current orbits \cp{Hansen2013,Chiang2013}?

In an attempt to address these questions, a great deal of effort has been invested in acquiring precise masses for a large number of these transiting planets. In recent years this has generated a much fuller understanding of the mass-radius relation, especially for sub-Neptune and super-Earth sized planets \cp{Weiss2013}. In particular, there are now several multi-planet {\it Kepler} systems like Kepler-11 with masses determined from Transit Timing Variations (TTVs) \cp[e.g][]{Lissauer2011a,Carter2012,Cochran2011,Lissauer2013}. Although rare, such systems are incredibly valuable because with both a mass and a radius we can estimate a planet's bulk composition using models of interior structure and thermal evolution \cp[e.g.][]{Rogers2010a, Nettelmann2011, Miller2011, Lopez2012, Valencia2013}. Thus far efforts have been focused on individually determining compositions for this handful of planets. This paucity stands in stark contrast to the over 3500 Kepler Candidates with only measured radii. Unfortunately the vast majority of these candidates are in dynamically inactive systems without strong TTVs or around distant stars too faint for radial velocity measurements.

Moreover, even with precise masses and radii there are inherent degeneracies which limit one's ability to constrain the bulk compositions of super-Earth sized planets. For ~1-2 $R_{\mathrm{\oplus}}$ planets the densities of water, silicate rocks, and iron (i.e. $\sim$ 1-10 $\mathrm{g \, cm^{-3}}$) are similar enough that it is impossible to uniquely constrain the relative abundance of these components \cp{Valencia2007,Rogers2010a}. To some extent models of planet collisions can set upper limits on the maximum iron or water mass fractions that are physically achievable \cp{Marcus2009,Marcus2010}, but for a given planet this still allows a wide range of internal compositions. Fortunately, models are still able to set clear and useful constraints on composition. In particular, thermal evolution models can set robust constraints on the fraction of a planet's mass in a H/He envelope. Due to its significantly lower density, even a relatively minor amount H/He (e.g., $\sim$1\% of total planet mass) has a large impact on planetary radius. For sub-Neptune sized planets $\sim$ 3-4 $R_{\mathrm{\oplus}}$, such an envelope will dominate a planet's size regardless of the abundance of other elements.

Moreover, for sub-Neptune sized planets at fixed bulk-composition, theoretical mass-radius curves are remarkably flat; i.e., planets with a given H/He abundance have very similar sizes regardless of their mass \cp{Lopez2012}. As a result, there is a remarkably tight relationship between planetary radius and H/He envelope fraction that is independent of planet mass. Critically, this opens up the hope of constraining compositions for the vast population of Neptune and sub-Neptune sized {\it Kepler} candidates without measured masses.  This is what we begin to explore in this paper.

Whenever possible it is still preferable to obtain a well measured mass. Planet mass is critical for understanding how volatile rich planets accrete their initial H/He envelope \cp{Bodenheimer2000,Ikoma2012} and whether they can retain it against X-ray and EUV driven photo-evaporation \cp{Lopez2012,Lopez2013,Owen2012,Owen2013}. Nevertheless, for systems of sub-Neptunes like Kepler-11, even factor of $\sim$2 uncertainties on planet masses are sufficient to tightly constrain composition with precise radii \cp{Lissauer2013}. This fact means that instead of only examining the {\it radius} distribution of {\it Kepler} candidates, we can begin thinking about a {\it composition} distribution. 
 
\section{Models}
\label{ModelSec}

In order to understand how planetary radius relates to planet mass and composition, it is necessary to fully model how a planet cools and contracts due to thermal evolution. For this work, we have used the thermal evolution presented in \ct{Lopez2012}, where additional model details can be found. Similar models are frequently used to track the evolution of sub-Neptunes and hot Jupiters. \cp[e.g,][]{Miller2011,Nettelmann2011}. Unlike \ct{Lopez2012} and \ct{Lopez2013}, here we do not consider the effects of photo-evaporation. Although photo-evaporation can have a large impact on the composition of a planet \cp[e.g.][]{Baraffe2006,Hubbard2007b,Lopez2012,Owen2012}, the effect on the thermal state of the interior is relatively minor \cp{Lopez2013}. Here we are primarily interested in the relationship between radius and composition as controlled by thermal evolution, as a result the effects of photo-evaporation can be ignored. In essence, present-day composition determines the radius, but that composition may have been strongly effected by formation and photo-evaporation.

At a given age, a model is defined by the mass of its heavy element core, the mass of its H/He envelope, the amount of incident radiation it receives, and the internal specific entropy of its H/He envelope.  As a default model, we assume an isothermal rock/iron core with an Earth-like 2:1 rock/iron ratio, using the ANEOS olivine \cp{Thompson1990} and SESAME 2140 Fe \cp{Lyon1992} equations of state (EOS). When determining composition error bars for observed planets, however, we varied this iron fraction from pure rock to the maximum possible iron fraction from impact models in \ct{Marcus2010}. For the H/He envelope we assume a fully adiabatic interior using the \ct{Saumon1995} EOS. In addition we consider the possibility of water-worlds and three component models using the H2O-REOS for water \cp{Nettelmann2008}. Finally atop the H/He envelope is a relatively small radiative atmosphere, which we assume is isothermal at the equilibrium temperature. We define a planet's radius at 20 mbar, appropriate for the slant viewing geometry in optical transits \cp{Hubbard2001}.  

In order to quantitatively evaluate the cooling and contraction of the H/He envelope, we use a model atmosphere grid over a range of surface gravities and intrinsic fluxes. These grids relate the surface gravity and internal specific entropy to the intrinsic flux emitted for a given model. These one-dimensional radiative-convective models are computed for solar metallicity and for 50$\times$ solar metallicity enhanced opacity atmospheres using the methods described in \ct{Fortney2007} and \ct{Nettelmann2011}. These atmosphere models are fully non-gray, i.e. wavelength dependent radiative transfer is performed rather than simply assuming a single infrared opacity. The atmospheres of Neptune and sub-Neptune sized planets might be significantly enhanced in metals \cp{Fortney2013} or host extended clouds that greatly enhance atmospheric opacity \cp{Morley2013}. Therefore, our two atmosphere grids are a way to make a simplified first estimate of the role of enhanced opacity on planetary thermal evolution. For all runs we use the H/He \ct{Saumon1995} EOS for the envelope.

At very early times and very low masses, the models reach gravities beyond the edge of our cooling grid. In such cases we logarithmically extrapolate the intrinsic temperature $T_{\mathrm{int}}$ as a function of gravity. This does not significantly affect our results, however, as the dependence of $T_{\mathrm{int}}$ on gravity is slight and the models are only at such low gravities in the first few Myr. 

Finally, we include heating from radioactive decay in the rock/iron core and the delay in cooling due to the core's heat capacity.  In order to correctly determine the mass-radius-composition relationship, it is vital to include these thermal evolution effects, since these will significantly delay cooling and contraction, particularly for planets less than $\sim$5 $M_{\mathrm{\oplus}}$. 

As with previous models, we assume that planets initially form with a large initial entropy according to the traditional "Hot-Start" model \cp{Fortney2007,Marley2007}. Specifically we start our models at an age of 1 Myr with a large initial entropy of 10 $k_{\mathrm{b}} \, \mathrm{baryon}^{-1}$. This assumption does not significantly affect any of our results since hot-start and cold-start models are indistinguishable by the time planets are $\sim$100 Myr old \cp{Marley2007,Lopez2012}. Moreover, \ct{Mordasini2013} recently showed that for planets less massive than Jupiter gravitional heating due to settling of heavy elements in the envelope can erase any difference between hot and cold starts.

For low-mass planets, the hot-start assumption results in extremely large initial radii $\gtrsim$10 $R_{\mathrm{\oplus}}$. However, as we explore in Section 3.2, such models cool extremely rapidly such that significant contraction has already occurred by several Myr. In general we present results at ages $>$10 Myr, when our results are insensitive to the initial choice of entropy.

\section{A Mass Radius Parameter Study}
\label{studysec}

\begin{figure}[h!] 
  \begin{center}
    \includegraphics[width=3.5in,height=6.0in]{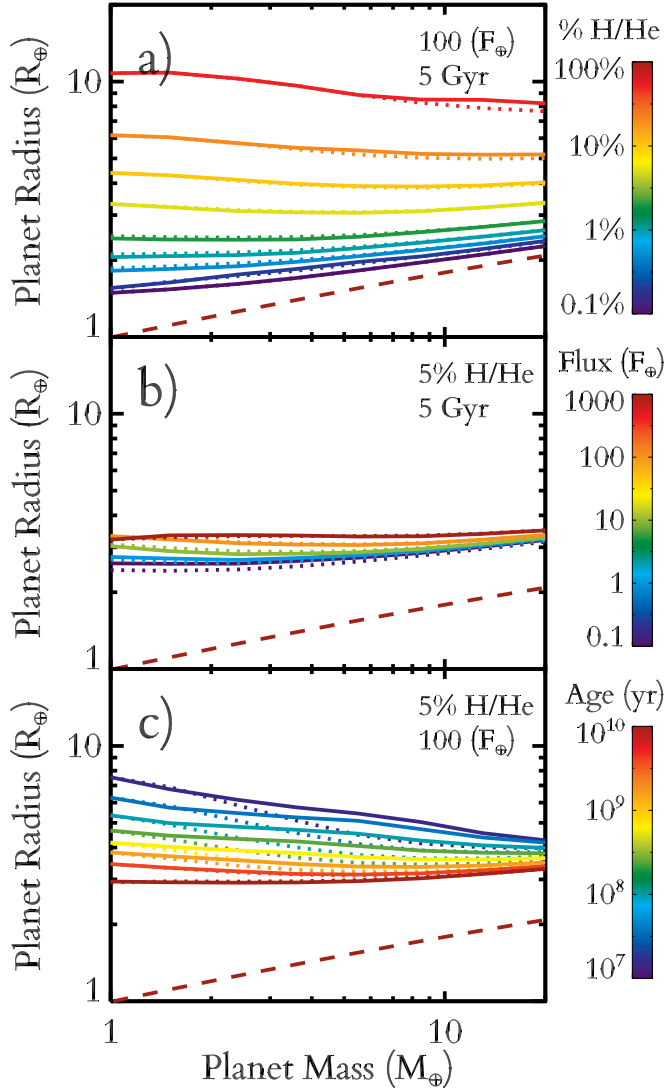}
  \end{center}
  \caption{Here we show model mass-radius relations from 1-20 $M_{\mathrm{\oplus}}$ and how these depend on composition, irradiation, and age, indicated by the colors. Solid lines correspond to enhanced opacity models, while dotted lines correspond to solar metallicity. The dashed rust-colored lines show the size of bare rocky planets with Earth-like compositions. Our default model is 5\% H/He, 5 Gyr old, and receives $\sim$100 $F_{\mathrm{\oplus}}$. In panel a) we vary the envelope fraction from 0.1-60\% H/He, this has by far the largest impact on planet size. Below $\sim$3\% H/He radius increases modestly with mass due to the dominance of the rocky core. For larger envelopes, the mass-radius relation is remarkably flat until for gas giant sized planets it decreases slightly with higher mass due to the increasing self gravity of the envelope. In panel b) we vary the incident flux a planet receives from 1-1000 $F_{\mathrm{\oplus}}$. Despite varying the irradiation by 4 orders of magnitude, the radius never changes by more than $\sim$30\%. Finally, in panel c) we show a time evolution from 10 Myr to 10 Gyr. At early times low-mass planets are larger than higher-mass planets due to their lower gravities. However, these low-mass planets are able to cool more rapidly which gradually flattens the mass-radius relation. \label{studyfig}}
\end{figure}

Planetary radius is an invaluable tool in understanding the nature of low-mass planets; however, without the aid of thermal evolution models like those used here, it can be quite difficult to interpret. In order to better understand the information contained in planet radii, we performed a detailed parameter study of our thermal evolution and structure models for sub-Neptune type planets with rock/iron cores and thick H/He envelopes.

As part of this parameter study we ran over 1300 thermal evolution models varying planet mass, incident flux, envelope fraction, and atmospheric metallicity. We covered planets from 1-20 $M_{\mathrm{\oplus}}$, 0.1-1000  $F_{\mathrm{\oplus}}$, 0.01-60\% H/He, for both solar metallicity and enhanced opacity models.We then recorded planet radius at every age from 10 Myr to 10 Gyr. The results of this study are summarized in Figure \ref{studyfig} and Tables 2-7. %\ref{sol100tab}-\ref{sol10tab50}. %In addition, we have made the full results of the parameter study available in the online supplemental tables.

Examining Figure \ref{studyfig}, it is immediately clear that iso-composition mass-radius curves are in fact remarkably flat for sub-Neptune or larger planets, at least once they are a few Gyr old. In each panel, we show theoretical mass-radius curves while varying the envelope fraction, incident flux, and age of the model planets. For the parameters that are not varying in each panel, we use representative values of 5\% H/He, 100 $F_{\mathrm{\oplus}}$, and 5 Gyr. 

Turning to panel a), we see the enormous effect that varying the H/He envelope fraction has on planetary radius. By comparison, any other changes to incident flux, age, or internal structure are secondary. For planets with envelopes $\sim$0.1\% of their total mass, the mass-radius curve does increase slightly from $\sim$1.5 $R_{\mathrm{\oplus}}$ at 1 $M_{\mathrm{\oplus}}$ to $\sim$2.5 $R_{\mathrm{\oplus}}$ at 20 $M_{\mathrm{\oplus}}$. For envelopes this insubstantial, a planet's size is still dominated by its rocky/iron core and so the mass-radius curves have a similar slope to the bare rock curve shown in Figure \ref{studyfig}. However, as we increase the envelope fraction, the mass-radius curves rapidly flatten, beginning at low-masses, until by $\sim$3\% H/He, the curves are almost completely flat.

By comparison, panel b) in Figure \ref{studyfig} shows the much more modest effect of varying the incident flux. More irradiated planets tend to be slightly larger because they have a large scale height in their atmospheres and because the irradiation alters the radiative transfer through their atmosphere, slowing their contraction \cp{Fortney2007}. Nonetheless, despite varying the incident flux by four orders of mangitude, planet radii vary by less $\sim$30\%. 

Finally, panel c) shows how these mass-radius curves evolve over time. At early times lower mass planets are significantly larger than higher mass planets due to their similarly large internal energies and lower gravities. Over time, however, these low mass planets are able to cool more rapidly than their more massive relatives, which gradually flattens the mass-radius curves. By the time the planets are $\sim$ 1 Gyr old we see the characteristically flat mass-radius curves for H/He rich planets.

\begin{figure}[h!] 
  \begin{center}
    \includegraphics[width=3.0in,height=8.3in]{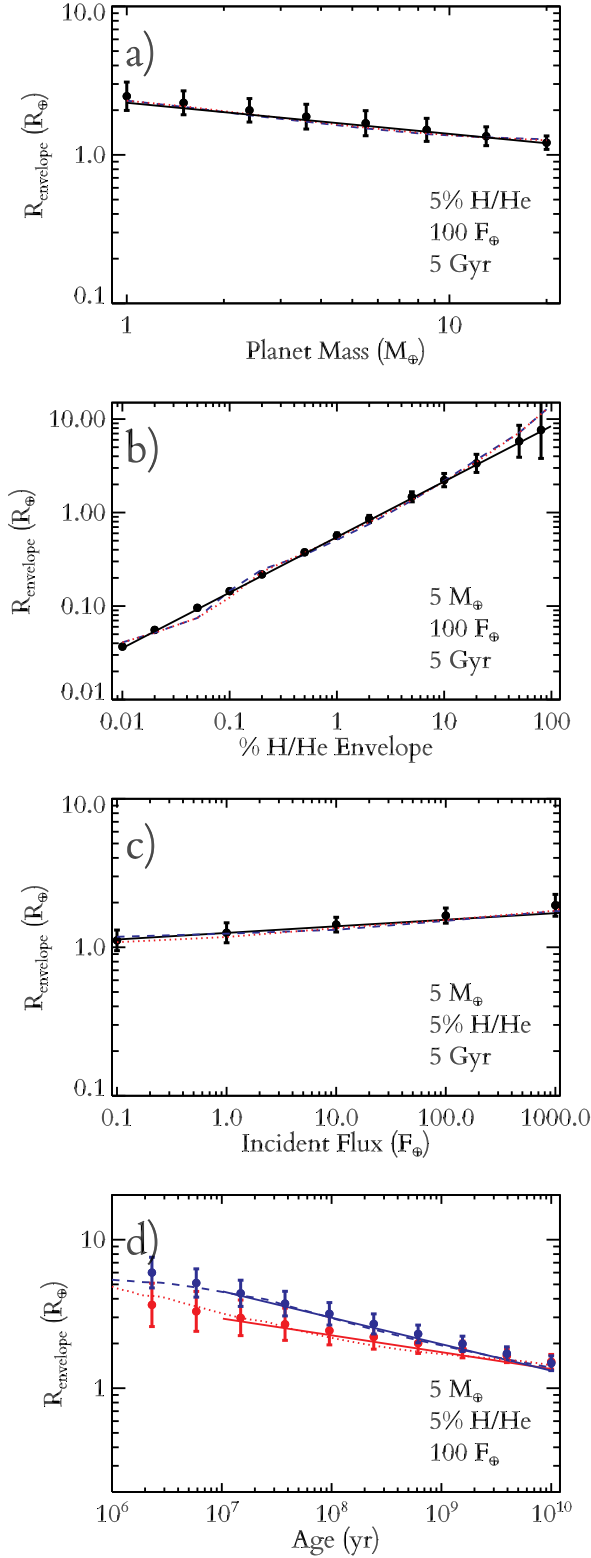}
  \end{center}
  \caption{Four panels showing how the radius of the H/He envelope $R_{\mathrm{env}}=R_{\mathrm{p}}-R_{\mathrm{core}}-R_{\mathrm{atm}}$ varies with planet mass, envelope mass fraction, incident flux, and planet age for representative values. Red dotted lines correspond to solar metallicity atmospheres, while blue dashed lines correspond to enhanced opacity. Solid lines indicate power-law fits as described in equation (\ref{powerlaweq}). Here we use default values of 5 $M_{\mathrm{\oplus}}$, 100 $F_{\mathrm{\oplus}}$, 5\% H/He, and 5 Gyr. \label{powerfig}}
\end{figure}

\subsection{Describing Radius with Power-Laws}
\label{powersec}

A quick inspection of Figure \ref{studyfig} makes clear that not all of a planet's properties have an equal impact on planet size. Planet mass and incident flux have only a modest impact on planet size, while planet age has a larger impact, particularly at younger ages. However, by far the largest determinate of a planet's size is the fraction of its mass in a H/He envelope. One way to quantify the relative importance of composition is to construct analytic fits for radius as a function of planet mass $M_{\mathrm{p}}$, envelope fraction $f_{\mathrm{env}}$, incident flux $F_{\mathrm{\oplus}}$, and age. In \ct{Lopez2013} we performed a similar analysis examining planets' vulnerability to photo-evaporative mass loss. 

Fortunately, the relationships between radius and each of these parameters are all reasonably well described by power-laws and the effects of each variable are relatively independent. As a result, we can do a reasonably good job of describing the results of our full parameter study with a set of four independent power-laws. The one caveat is that we do not fit for the total planet radius $R_{\mathrm{p}}$, but instead the radius of the H/He envelope $R_{\mathrm{env}} \approx R_{\mathrm{p}}-R_{\mathrm{core}}$, where $R_{\mathrm{core}}$ is the size of the rock/iron core. We do this because as $f_{\mathrm{env}}$ approaces zero, the planet radius does not approach zero but instead assymptotes to $R_{\mathrm{core}}$. 

To first order, however, the rock/iron equation of state is very incompressible and so we can approximate $R_{\mathrm{core}}$ with the mass-radius curve of a envelope free rocky planet. Assuming an Earth-like composition, then $R_{\mathrm{core}}$ is described by equation (\ref{rockeq}) to within $\sim$2\%. If we also allow the iron-fraction of the core to vary then this error rises to $\sim$10\%, but for the qualitative analysis we attempting here such errors are unimportant. $M_{\mathrm{core}}$ in equation (\ref{rockeq}) refers to the mass of the rock/iron core, which for sub-Neptune sized planets is approximately the same as the total planet mass $M_{\mathrm{p}}$.

\begin{equation}\label{rockeq}
  R_{\mathrm{core}} = \left(\frac{M_{\mathrm{core}}}{M_{\mathrm{\oplus}}}\right)^{0.25} \approx \left(\frac{M_{\mathrm{p}}}{M_{\mathrm{\oplus}}}\right)^{0.25}
\end{equation}

Likewise, we must make a small correction to account for the size of the radiative upper atmosphere. To first approximation, this atmosphere is isothermal at the planet's equilibrium temperature $T_{\mathrm{eq}}$. For sub-Neptune sized planets at several Gyr, the radiative-convective boundary is typically $\sim$100-1000 bar. For transiting planets the broadband optical radius is typically $\sim$20 mbar, or $\approx$8-10 scale heights higher. Thus the size of the radiative atmopshere is approximately given by equation (\ref{atmeq}), where $g$ is a planet's gravity and $\mu_{\mathrm{H/He}}$ is the mean molecular weight. Generally however, this correction is typically quite small $\sim$0.1 $R_{\mathrm{\oplus}}$ except at the very highest levels of irratiation.

\begin{equation}\label{atmeq}
  R_{\mathrm{atm}} \approx \log{\left(\frac{100 \, \mathrm{bar}}{20 \, \mathrm{mbar}}\right)} H \approx 9 \left(\frac{k_{\mathrm{b}} \, T_{\mathrm{eq}} }{g \, \mu_{\mathrm{H/He}}}\right)
\end{equation}

With equations (\ref{rockeq}) and (\ref{atmeq}) in place, we can now fit for $R_{\mathrm{env}}$, and then simply add $R_{\mathrm{core}}$ and $R_{\mathrm{atm}}$to get the total radius. The results of these fits are summarized in Figure \ref{powerfig} and equation (\ref{powerlaweq}). Figure \ref{powerfig} compares our power-law fits to the results of our full models for representative values of $M_{\mathrm{p}}$, $f_{\mathrm{env}}$, $F_{\mathrm{\oplus}}$, and age. The error bars in each panel show the 1$\sigma$ scatter about the power-law fits for the full suite of models in our parameter study. Remarkably, this simple power-law description does a reasonable job of reproducing the results of our full model. In general, the analytic formulation in equation (\ref{powerlaweq}) matches our full models to within $\sim$0.1 dex.

For the age evolution, we fit separate power-laws for solar metallicity and enhanced opacity models. The solar metallicity models cool more rapidly initially. As a result, they are already relatively cold by $\sim$100 Myr and so the subsequent contraction is slower. However, the enhanced opacity models must eventually cool and by several Gyr any differences are erased. We fit power-laws only to the evolution after 100 Myr. For solar metallicity $R_{\mathrm{env}} \sim t^{0.11}$ while for enhanced opacity $R_{\mathrm{env}} \sim t^{0.18}$. Equation (\ref{powerlaweq}) shows the results for the enhanced opacity models.

\begin{equation}\label{powerlaweq}
\begin{split}
 R_{\mathrm{env}} = R_{\mathrm{p}}-R_{\mathrm{core}}-R_{\mathrm{atm}} = 2.06 \, R_{\mathrm{\oplus}} \left(\frac{M_{\mathrm{p}}}{M_{\mathrm{\oplus}}}\right)^{-0.21}
\\
\times \left(\frac{f_{\mathrm{env}}}{5\%}\right)^{0.59} \left(\frac{F_{\mathrm{p}}}{F_{\mathrm{\oplus}}}\right)^{0.044} \left(\frac{age}{5 \, \mathrm{Gyr}}\right)^{-0.18}
\end{split}
\end{equation}
%% add separate fit for solar metallicity

It is important to note however, that the results of these fits are only meant to be a rough approximation of the full models summarized Figure \ref{studyfig} and tables 2-6. These fits are done purely to help understand the qualitative behavior of our thermal evolution models, not to be used in place of the full models. Also, equation (\ref{powerlaweq}) shows the fit to our the enhanced opacity models. At late times the solar metallicity models have a slightly shallower dependence on age, due to more rapid cooling at early ages.

Nonetheless, equations (\ref{rockeq}) and (\ref{powerlaweq}) do make several things quite clear. First of all, we can now quantify the importance of H/He envelope fraction; doubling $f_{\mathrm{env}}$ has an order of magnitude larger effect on $R_{\mathrm{p}}$ than doubling $F_{\mathrm{p}}$ and more than twice as large as an effect of doubling the age. We can also now see how flat the mass-radius curves are. Although, $R_{\mathrm{env}}$ decreases slightly with mass, this is almost exactly balanced by the increase in $R_{\mathrm{core}}$ with increasing mass. So long as $R_{\mathrm{env}} \gtrsim R_{\mathrm{core}}$, then these terms will roughly balance and the mass-radius curves will be quite flat. This typically happens for planets that are $\gtrsim$1\% H/He or $\gtrsim$2.5 $R_{\mathrm{\oplus}}$. Thus for most of {\it Kepler}'s Neptune and sub-Neptune sized planets, radius is quite independent of planet mass and is instead a direct measure of bulk H/He envelope fraction.

\begin{figure}[h!] 
  \begin{center}
    \includegraphics[width=3.5in,height=2.5in]{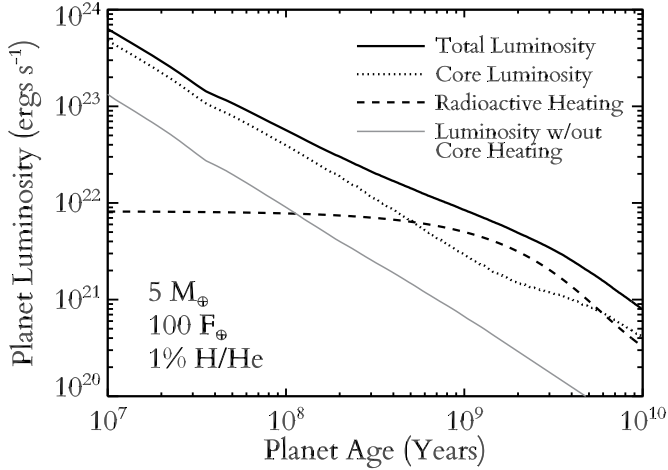}
  \end{center}
  \caption{Here we show the planet luminosity budget vs. time for a representative example thermal evolution model with 1\% H/He on a 5 $M_{\mathrm{\oplus}}$ planet, receiving 100 $F_{\mathrm{\oplus}}$ from a sun-like star. The black solid line shows the overall cooling rate while the dotted and dashed lines show the cooling rate of the rock/iron core and the heating from radioactive decay, respectively. The solid gray line shows the cooling rate if we ignore radioactivity and the need to cool the core. This clearly demonstrates the need to include these terms when calculating the thermal evolution of sub-Neptune like planets.\label{evolfig}}
\end{figure}

\begin{figure}[h!] 
  \begin{center}
    \includegraphics[width=3.5in,height=2.5in]{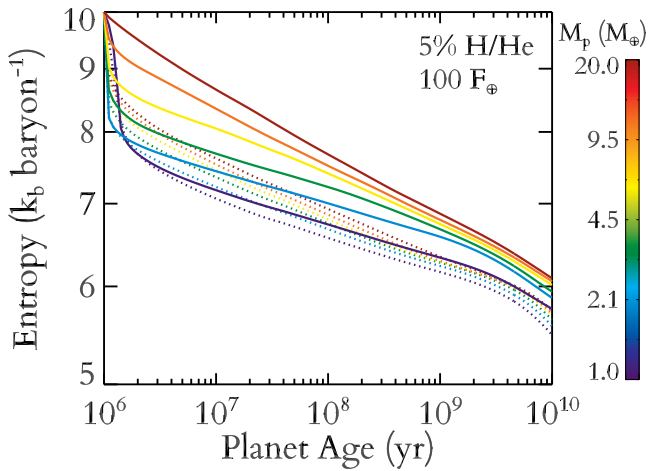}
  \end{center}
  \caption{Shown is an example calculation in which all models start at the same young age and initial specific entropy.  Internal specific entropy in the H/He envelope vs. time is shown for various planet masses. Solid lines show enhanced opacity, while dotted show solar metallicity. Planets start with large initial entropy, then rapidly cool. By 10-100 Myr, the models are insensitive to the choice of initial entropy. Low-mass planets experience more rapid cooling, leading to the flat mass-radius curves seen in Figure \ref{studyfig}. Solar metallicity models cool rapidly at young ages and then experience more gradual cooling, while enhanced opacity models cool more steadily at all ages.\label{entfig}}
\end{figure}

\begin{figure}[h!] 
  \begin{center}
    \includegraphics[width=3.5in,height=2.5in]{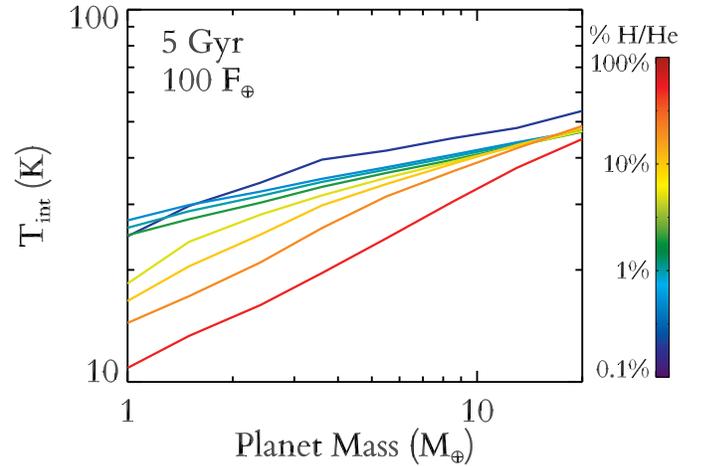}
  \end{center}
  \caption{Intrinsic temperature $T_{\mathrm{int}}$, i.e., the equivalent blackbody temperature a planet's net outgoing flux, vs. planet mass for 5 Gyr old planets receiving 100 $F_{\mathrm{\oplus}}$ with enhanced opacity atmospheres. Colors show different H/He envelope fractions. Clearly, by several Gyr lower-mass planets are signficantly colder than higher mass planets. This demonstrates the need to perform full thermal evolution calculations. Simply assuming a fixed luminosity per mass will greatly overestimate the size of planets below $\sim5$ $M_{\mathrm{\oplus}}$. \label{tintfig}}
\end{figure}

\subsection{Why is the Mass-Radius Relation Flat?}
\label{flatsec}

One of the key features of our thermal evolution and structure models is the relative flatness of mass-radius curves at fixed H/He envelope fraction. In sections \ref{studysec} and \ref{powersec}, we showed that for planet with $\gtrsim$1\% H/He, planet size is more or less indepent of mass. Thus far, however, we have not explained the origin of this flatness.

In fact, searching through the literature will show a wide range of mass-radius curves with very different behavior at low masses. Although, all the models tend to agree above $\sim$10-20 $M_{\mathrm{\oplus}}$, there can be large disagreements below $\sim$5 $M_{\mathrm{\oplus}}$. In some cases, radius decreases with decreasing mass in much the same way as the Earth-like mass radius curves in Figure \ref{studyfig}. In other cases, the radius increases to implausibly large sizes due to the planets' lower gravity \cp{Rogers2011}. Generally, these models face one of two limitations. Either they they ignore the contributions of the rock/iron core to the thermal evolution, i.e., the need to cool the core and heating from radioactive decay, or they do not perform an evolution calculation at all and instead use static structure models in which the internal energy of the planet is treated as a free parameter.

For the Neptune and sub-Neptune sized planets that we are focusing on here, $\sim$90-99\% of a planet's mass is contained in the rock/iron core. As a result, ignoring the effects of that core on the thermal evolution will significantly underestimate a planet's cooling timescale, and therefore its radius. This is a common simplification with thermal evolution models, like our own, that were originally developed to model massive gas giants, where the core has a negligible impact on the overall thermal evolution. The importance of these effects, however, is clearly demonstrated in Figure \ref{evolfig}, which shows the various contributions to the overall thermal evolution, for a typical 5 $M_{\mathrm{\oplus}}$, 1\% H/He sub-Neptune sized planet. At every age, the cooling luminosity of the planet is dominated by these core cooling and heating terms. At early times, the thermal evolution is largely regulated by the need to cool the rock/iron core with its relatively large heat capacity \cp{Alfe2002,Guillot1995}. At ages $\gtrsim$1 Gyr, radioactive heating also becomes comparable to the core cooling rate, thanks mostly due to the decay of $^{40}$K \cp{Anders1989}. On the other hand, ignoring these terms leads to a planet that is $\sim$30-100$\times$ less luminous at late times, and underestimates the final radius by $\sim$0.5 $R_{\mathrm{\oplus}}$. Some models \cp[e.g.,][]{Mordasini2012c}, make the compromise of including radiogenic heating but not including the effect of the core's heat capacity. This is much better than ignoring the core altogether, but as shown in Figure \ref{evolfig} both terms are important and this will lead to underestimating the radii of sub-Neptune planets, especially at ages $\lesssim$1 Gyr.

On the other hand, it is also quite common to use static internal structure models which do not track a planet's thermal evolution, but instead assume a fixed specific luminosity (i.e. power per unit mass), which is then treated as a free variable \cp{Rogers2011}. This is a common simplification made when a small H/He envelope is added to detailed models of terrestrial planets, for which the cooling history is harder to determine and has little impact on overall planet size \cp{Valencia2007}. When calculating possible compositions for a single planet \cp[e.g.,][]{Rogers2010b}, this is fine, so long as the resulting uncertainty in the internal energy is accounted for. However, when plotting iso-composition mass-radius curves, this leads to an unphysical upturn at low masses. Low-mass planets of course have lower surface gravities, so assigning them the same specific luminosities as more massive planets will significantly inflate their radii.

In reality though, low mass planets are able to cool much more quickly. Partly this is due to their low gravities which slightly increases the rate of radiative transfer through their atmospheres \cp{Fortney2007}. Mostly, however, it is simply due to the fact that lower mass planets have similar radiating surface areas to slightly higher mass planets, but significantly smaller total internal energies. As a result, even if different mass planets start with similar specific internal energies, low mass planets will more quickly deplete their thermal energy reserves, leading to much shorter cooling times. 

This fact is summarized in Figures \ref{entfig} and \ref{tintfig}. Figure \ref{entfig} shows various cooling curves for the internal entropy in the H/He envelope. Planets start with large initial entropy, and therefore radii. Models rapidly cool for the first few Myr until the cooling timescale is comparable to the age. As described above, all things being equal, less massive planets will tend to have shorter cooling timescales due to their smaller energy reservoirs. As a result, lower mass models tend to be colder at all ages. This counter balances the fact that lower mass planets have lower gravities and produces the flat mass-radius curves seen in Figure \ref{studyfig}. This result is insensitive to our choice of initial entropy for ages $\gtrsim$10 Myr.

As in Figure, \ref{powerfig} solar metallicity models cool rapidly for their first $\sim$10 Myr and then contract more slowly. The enhanced opacity models on the other hand, cool more steadily throughout their history. Eventually however, the enhanced opacity models must also cool and contract and by several Gyr they have largely erased any differences with the solar models. At the same time, there is a slight change in the cooling rates due to the decay of $^{40}K$.

Figure \ref{tintfig} shows the end result of this evolution. Here we show planetary intrinsic temperature $T_{\mathrm{int}}$ versus planet mass for various H/He envelope fractions for 5 Gyr old planets receiving 100 $F_{\mathrm{\oplus}}$. $T_{\mathrm{int}}$ is the equivalent blackbody temperature of the net radiation leaving a planet, effectively it is the temperature the planet would have if the parent star was removed. As we can see, by 5 Gyr, low mass planets are always significantly cooler than higher mass planets at the same compositions, regardless of H/He fraction or atmosphere metallicity. The result is that the lower gravities of lower mass planets is balanced out by their shorter cooling timescales and we arrive at the flat mass radius curves shown in Figure \ref{studyfig}.

\section{The Mass-Composition Relation}
\label{compsec}

\begin{figure}[h] 
  \begin{center}
    \includegraphics[width=3.5in,height=2.5in]{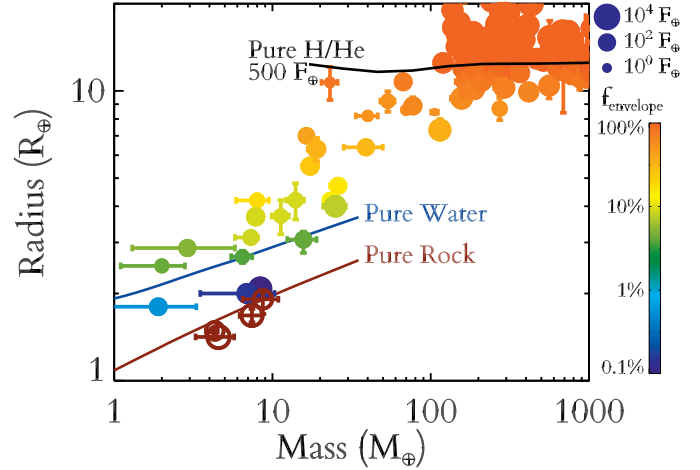}
  \end{center}
  \caption{Planetary radius vs. mass for all ~200 transiting planets with measured masses. Each planet is colored according to the fraction of its mass in a H/He envelope, assuming a water-free interior. Rust-colored open circles indicate potentially rocky planets. Points are sized according to the incident flux they receive from their parent stars, relative to $F_{\mathrm{\oplus}}$ the flux that the Earth receives from the Sun. For comparison, we include theoretical mass-radius relations for pure silicate rock, pure water, and pure H/He at 500 $F_{\mathrm{\oplus}}$. There is a very strong correlation between planetary radius and H/He envelope fraction, both of which are more weakly correlated with mass up to $\sim$100 $M_{\mathrm{\oplus}}$. \label{mrfig}}
\end{figure}

Using our thermal evolution and structure models, we calculated H/He envelope fractions for all $\sim$200 confirmed planets with well determined masses, assuming a water-free interior. We excluded any planets which only have upper limits on mass or purely theoretical mass constraints. We used masses and radii from exoplanets.org \cp{Wright2011}, except for where there are more recent values in the literature. For CoRoT-7b, the five inner Kepler-11 planets, and 55 Cancri e we used masses and radii from \ct{Hatzes2011}, \ct{Lissauer2013}, and \ct{Dragomir2013b}, respectively. We exclude confirmed planets with analytical TTV mass estimates from \ct{Xie2012} due to the degeneracy between planet mass and free eccentricity. For inflated hot Jupiters with radii larger than that of pure H/He, we simply assigned 100\% H/He since such planets are beyond the scope of this work. Meanwhile, for potentially rocky planets like CoRoT-7b \cp{Leger2009,Queloz2009} and Kepler-10b \cp{Batalha2011}, we set strict upper limits on the size of any potential H/He envelope. Table 1 summarizes the results for 33 planets with measured masses $<100 \, M_{\mathrm{\oplus}}$ and radii $<12 \, R_{\mathrm{\oplus}}$.

In order to calculate the uncertainty on these compositions we included the affects of 1$\sigma$ variations in the observed planet masses, radii, ages, and levels of irradiation. In addition, we included theoretical uncertainties on core iron fraction, core heat capacity, atmospheric albedo, etc., as described in \ct{Lopez2012}. In general, uncertainties in the stellar radius and therefore the planetary radius are the dominant source of uncertainty. Typically this is followed by the unknown iron fraction in the core which is typically equivalent to a $0.1$ $R_{\mathrm{\oplus}}$ uncertainty in the radius for low-mass planets. 

Figure \ref{mrfig} plots the current measured mass-radius relation with 1$\sigma$ uncertainties for all confirmed transiting planets with measured masses up to $1000 \, M_{\mathrm{\oplus}}$ and radii $20 \, R_{\mathrm{\oplus}}$. The color of each point shows the H/He envelope fractions calculated by our models. Rust-colored open circles show potentially volatile-free rocky planets. Meanwhile, the size of the points correspond to the incident flux that each planet receives from its parent star, relative to $F_{\mathrm{\oplus}}$, the incident flux that the Earth receives from the Sun.

Finally, we include three theoretical iso-composition curves. The rust colored curve shows pure silicate rock (specifically olivine). The dark blue curve corresponds to pure water worlds on a 10 day orbit around a 5 Gyr old Sun-like star, however, varying these details does not significantly change the curve. Finally, the black curve corresponds to pure H/He hot Jupiters receiving 500 $F_{\mathrm{\oplus}}$ (i.e., 500 times the current incident flux that the Earth receives from the Sun) from a 5 Gyr old Sun-like star. Roughly speaking, this last curve forms the dividing line between the inflated and non-inflated hot Jupiters.

Several features of the mass-radius relation are immediately apparent. As noted in \ct{Weiss2013}, there is a roughly power-law increase in radius from $\sim$ 1-100 $M_{\mathrm{\oplus}}$, above which radius saturates at approximately a Jupiter radius. Below $\sim$ 10 $M_{\mathrm{\oplus}}$ there is a particularly large scatter in radius, with planets ranging from the potentially rocky to sub-Neptune sized planets with $\sim$3\% H/He. For low-mass planets there is also an inverse correlation between radius and incident flux which may be due to photo-evaporative loss of H/He \cp{Lopez2012, Owen2013}. 

Above $\sim 100 M_{\mathrm{\oplus}}$ we find the true gas giants including the highly inflated hot Jupiters. Here the correlation with incident flux is the reverse of that at low-mass with the most irradiated planets being extremely inflated. It is unclear why there do not appear to be any super-inflated hot Jupiters below $\sim 100 M_{\mathrm{\oplus}}$, it is possible that such planets would be unstable to photo-evaporation or Roche-lobe overflow \cp{Jackson2010} or have a high mass fraction of heavy elements \cp{Miller2011}.

Turning to the compositions of these planets, it is immediately clear that H/He envelope fraction is strongly correlated with both planet mass and radius. Below $\sim 10$ $M_{\mathrm{\oplus}}$, planets range from potentially rocky super-Earth sized planets to sub-Neptunes with a few percent H/He envelopes. From $\sim 10-50$ $M_{\mathrm{\oplus}}$, we have the Neptunes and super-Neptunes with $\sim10-30$\% of their mass in the envelope. Finally above $\sim 50$ $M_{\mathrm{\oplus}}$, planets transition to true gas giants where both the mass and radius are completely dominated by gas accreted during formation.

However, on closer inspection, where there is scatter in the mass-radius relationship it is the planet radius that correlates with composition. We argue here that planet radius is first and foremost a proxy for a planet's H/He inventory. The fact that both composition and radius correlate with mass is due to the fact that more massive planets are able accrete more gas during formation.

The radius saturates at $\sim$100 $M_{\mathrm{\oplus}}$ because planet size does not simply increase with increasing H/He mass but rather with increasing H/He mass {\it fraction}. As shown in section \ref{studysec}, there is an approximately power-law relationship between the size of a planet's H/He envelope and the planets H/He mass fraction. A 100 $M_{\mathrm{\oplus}}$ planet with a 10 $M_{\mathrm{\oplus}}$ core, is already 95\% H/He, as a result doubling the mass will not significantly increase the H/He envelope fraction or the radius.

Although incredibly valuable, planetary radius is in some sense not a fundamental parameter of a planet. It changes as a planet evolves and only through the aid of thermal evolution and structure models like those used here, does it tell us about a planet's structure and composition. Fortunately, however such models allow us to translate radius into an estimate of planet composition.
\begin{figure}[h!] 
  \begin{center}
    \includegraphics[width=3.5in,height=2.5in]{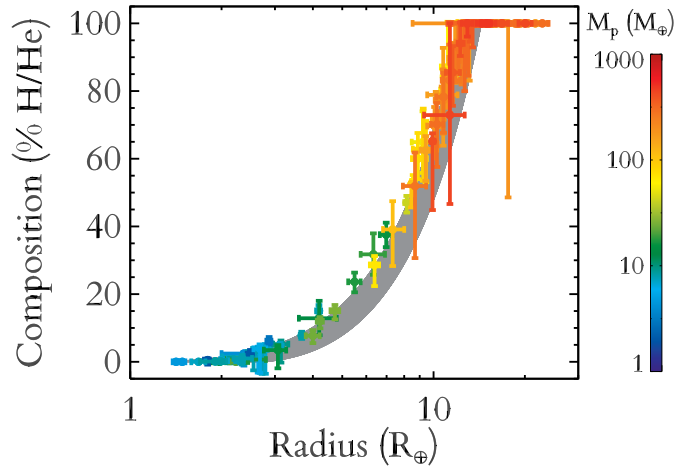}
  \end{center}
  \caption{H/He envelope fraction vs. planet radius, for the ~200 transiting planets shown in figure \ref{mrfig}. Here each planet is color-coded according to its mass. The grey shaded region shows the effect of varying the water abundance of the interior. Clearly there is a very tight correlation between size and envelope fraction, lending credence to our claim that radius can be used as a proxy for planetary composition.\label{rcfig}}
\end{figure}

\begin{figure}[h!] 
  \begin{center}
    \includegraphics[width=3.5in,height=2.5in]{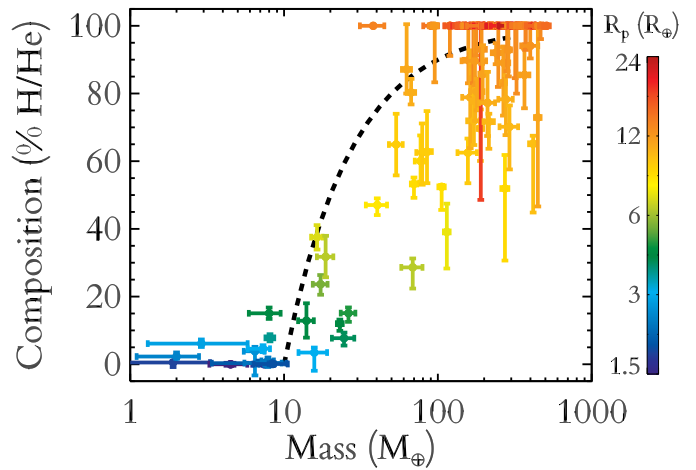}
  \end{center}
  \caption{Similar to figure \ref{rcfig} but with H/He envelope fraction plotted against planetary mass, and color-coded by radius. Below $\sim$10 $M_{\mathrm{\oplus}}$ there is a mix of rocky planets, possible water worlds, and sub-Neptunes with a few percent H/He. From $\sim$10-100 $M_{\mathrm{\oplus}}$ there is a strong increase in both radii and H/He envelope fraction transitioning from Neptune sized planets with $\sim$10\% H/He up to true gas giants that are almost entirely H/He. Above $\sim$100 $M_{\mathrm{\oplus}}$ we find the familiar hot Jupiters, many of which have large inflated radii. The dashed black line shows a toy-model in which all planets have a 10 $M_{\mathrm{\oplus}}$ core. \label{mcfig}}
\end{figure}

Figure \ref{rcfig} shows the observed sample of transiting planet except that here we have plotted H/He envelope fraction against radius. This clearly demonstrates the close relationship between the observed radius and the fundamental bulk composition. At a given radius, planet mass, shown by the color bar, can span up to a factor of $\sim$3. Nonetheless the scatter in envelope fraction is typically only $\sim$0.3 dex. This is what we mean when we state that radius is primarily a proxy for composition. 

Thus far, however, we have only considered dry interiors with H/He envelopes atop rock/iron cores. The gray shaded region in Figure \ref{rcfig} shows the effect of varying the water abundance of planets in our model. Using our three layer models we varied the water abundance of the interior from completely-dry, up to 90\% of core mass, where by ``core'' we mean the combined mass of the rock and water layers. For clarity, we then fit power-laws to best fit radii and compositions under both scenarios; the gray shaded region shows the area in between these fits. Clearly, allowing this degeneracy does slightly increase the scatter in the radius-composition relationship. Nonetheless, above $\sim$3 $R_{\mathrm{\oplus}}$ this does not alter the conclusion that radius and H/He envelope fraction are intimately related.

As a result, this means that we can recast the mass-radius relationship in Figure \ref{mrfig} as a mass-{\it composition} relationship. This is shown in Figure \ref{mcfig}. By doing this we have transformed the observable mass-radius relationship into one that is directly relatable to models of planet formation. Here we can clearly see that there is a fundamental change in the relationship around $\sim$10 $M_{\mathrm{\oplus}}$. Below this planets typically have less than $\sim$5\% of their mass in H/He with no clear relationship between envelope fraction and mass. Above this, however, we see a steady rise in envelope fraction from sub-Neptunes up to gas giants. 

These trends are all understandable in the light of the traditional core accretion model of planet formation \cp[e.g.,][]{Hayashi1985,Bodenheimer1986}. If a planet's rocky core becomes sufficiently massive, typically $\sim$5-10 $M_{\mathrm{\oplus}}$, then its gravity becomes sufficiently strong to trigger runaway accretion from the disk. For comparison, the dashed black line in Figure \ref{mcfig} shows a the simple toy model in which all planets have 10 $M_{\mathrm{\oplus}}$ core with solar metalcity H/He envelopes. In reality most planets lie to the right of this curve, possibly indicating that the accreted additional planetessimals embedded in the nebula \cp{Mordasini2013}. 

The results of Figures \ref{mrfig} and \ref{mcfig} are consistent with this traditional picture of core accretion. Below $\sim$10 $M_{\mathrm{\oplus}}$, according to our models planets are almost entirely composed of heavy elements by mass. Above this, most planets are roughly consistent with $\gtrsim 10$ $M_{\mathrm{\oplus}}$ of heavy elements along with accreted H/He envelopes. There is of course a simplified view of planet formation. In reality there is considerable variation in disk mass, lifetime, metallicity, planet history, etc. all of which introduces considerable scatter into the mass-composition relation. Nonetheless, Figure \ref{mcfig} offers clear evidence for the core-accretion model of planet formation, at least for the close-in planets found by {\it Kepler}

\section{Super-Earth vs. Sub-Neptune}

Throughout this paper, we have repeatedly used the terms super-Earth and sub-Neptune to refer to low-mass {\it Kepler} planets. What exactly is the difference between these classes of planets? For our purpose a sub-Neptune is any planet whose radius cannot be explained by a bare rock/iron model, i.e., it must have some sort of large optically-thick H/He or water envelope. Super-Earth on the other hand implies a more terrestrial planets, one that may have a solid or liquid surface and where the atmosphere, if any, contributes a negligible fraction to the planet's size. Although this may seem like semantics, one of the long-term goals of exoplanet science is to search for biomarkers in the transmission spectra of potentially habitable super-Earths. Whether or not a planet has a large H/He envelope tens of kbar deep has very important implications for habitability.

\begin{figure}[h!] 
  \begin{center}
    \includegraphics[width=3.5in,height=2.5in]{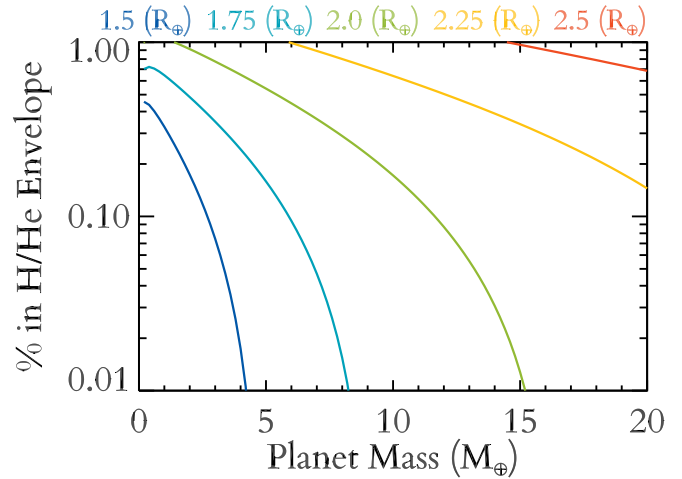}
  \end{center}
  \caption{H/He envelope fraction vs. planet mass for super-Earth and sub-Neptune sized planets. Curves are color-coded according to planet radius ranging from 1.5-2.5 $R_{\mathrm{\oplus}}$. Here we assume water-free sub-Neptunes with H/He envelopes atop Earth-like rocky cores.\label{minfig}}
\end{figure}

The current definition used by the {\it Kepler} mission is that planets 1.5-2.0 $R_{\mathrm{\oplus}}$ are super-Earths, while planets 2.0-4.0 $R_{\mathrm{\oplus}}$, are sub-Neptunes. These round numbers however, do not quite correspond to our more physically motivated definition of whether or not a planet has a thick envelope. Figure \ref{minfig} plots the minimum H/He envelope fractions required by our models vs. planet mass for several different radii in the 1.5-2.0 $R_{\mathrm{\oplus}}$ super-Earth/sub-Neptune transition region. 

%%%% Fix this!!!

It is quite difficult to construct a 2.0 $R_{\mathrm{\oplus}}$ planet that does not have some sort of thick envelope. Assuming an Earth-like interior, such planets would have to be 16.5 $M_{\mathrm{\oplus}}$, to explain their size without any type of envelope. For a completely iron-free interior, it is possible to construct a 2.0 $R_{\mathrm{\oplus}}$ that is only 11 $M_{\mathrm{\oplus}}$. However, completely iron-free is probably not a realistic composition for planets of several earth masses. Indeed both Kepler-10b and CoRoT-7b, may be slightly enhanced in iron compared to the Earth \cp{Batalha2011,Hatzes2011}. 

This stands in contrast to the observed sample of likely rocky planets all of which are $<$10 $M_{\mathrm{\oplus}}$. It is possible that more massive rocky planets are yet to be found, however, the ${\it Kepler}$ is essentially complete for 2.0 $R_{\mathrm{\oplus}}$ within 100 days \cp{Petigura2013}. For follow-up RV and TTV mass measurements to have missed a population of $<$10 $M_{\mathrm{\oplus}}$ rocky planets, they would need to somehow be biased against more massive and therefore easier to detect planets. Moreover, there are basic arguments in core-accretion theory that lead us to expect that there should not be $\sim$20 $M_{\mathrm{\oplus}}$ rocky planets. By the time a planet is $\sim$10 $M_{\mathrm{\oplus}}$, its gravity should be sufficiently strong that it should be able to accrete a substantial H/He envelope from the disk \cp{Ikoma2012}, and for periods $\gtrsim$10 days be able to retain it against photo-evaporation \cp{Lopez2013}.

On the other hand, if we assume a more typical low-mass planet with a 5 $M_{\mathrm{\oplus}}$ Earth-like core, then to be 2.0 $R_{\mathrm{\oplus}}$ it would need 0.5\% of its mass in a H/He envelope. This may not sound like much, but it corresponds to $\sim$20 kbars of hydrogen and helium, $\sim 20 \times$ higher than the pressure at the bottom of the Marianias Trench. Moreover, the temperature at the bottom of such an envelope would be $\gtrsim$3000 K, even for ages of several Gyr. We believe that such a planet is more properly classified as a sub-Neptune. As a result, 2.0 $R_{\mathrm{\oplus}}$ is more of a quite hard upper limit for the size of a envelope-free super-Earth and most of the planets between $\sim$ 1.75 and 2.0 $R_{\mathrm{\oplus}}$ are likely to be H/He rich sub-Neptunes.

If 2.0 $R_{\mathrm{\oplus}}$ is really the hard upper limit for the super-Earth/sub-Neptune transition, then what is the lower limit? As shown in Figure \ref{minfig}, for planets $\lesssim$1.5 $R_{\mathrm{\oplus}}$ it is entirely possible to explain their radii without any H/He. Moreover if such planets do have any H/He, then it must be $\lesssim$0.1\% of their mass, even if we assume a maximally iron-rich core. This is small enough of envelope that the rock/iron core dominates the planets size. Moreover, as shown in \ct{Lopez2013} and \ct{Owen2013}, such tenuous envelopes are quite-vulnerable to being completely photo-evaporated, at least at period $\lesssim$100 days. This does not exclude the possibility that 1.5 $R_{\mathrm{\oplus}}$ cannot have large water envelopes, but it does suggest that they are unlikely to have large H/He envelopes.

To summarize, we can say that 2.0 $R_{\mathrm{\oplus}}$ is likely a hard upper limit for the maximum size of envelope-free rocky super-Earths and 1.5 is likely a lower limit for the minimum size of a H/He rich sub-Neptune. As a result, we suggest using 1.75 $R_{\mathrm{\oplus}}$ rather than 2.0 $R_{\mathrm{\oplus}}$ for the dividing line between these classes of planets.

\section{Discussion}
In Sections \ref{studysec} and \ref{compsec}, we showed that planetary radius is to first order a proxy for a planet's composition above $\sim$2 $R_{\mathrm{\oplus}}$. This means that the observed radius occurrence distribution for {\it Kepler} candidates found by \ct{Fressin2013} and \ct{Petigura2013}, is in reality a {\it composition} occurrence distribution for close in planets at several Gyr. In particular, \ct{Fressin2013} and \ct{Petigura2013} found that there is a sharp, roughly power-law like drop off in the frequency of planet occurrence above $\sim$3  $R_{\mathrm{\oplus}}$, while below this there a plateau in the plant occurrence rate down to at least 1 $R_{\mathrm{\oplus}}$.

This distribution makes sense in the light of traditional core accretion theory. The timescale for planetesimal collisions to form rocky planets is short compared to the typical lifetime of a disk and such planetesimals are preferentially concentrated deep in the stars potential well, so nature easily makes large populations of irradiated rocky planets \cp{Chiang2013,Hansen2013}.

At larger sizes, planets are limited by their ability to accrete a H/He envelope from the disk before the disk dissipates \cp{Bodenheimer2000,Ikoma2012,Mordasini2012c}. In these models the accretion of the envelope is limited by the ability of the proto-planetary envelope to cool and contract. This makes it difficult to accrete larger initial H/He envelopes, particularly if the {\it Kepler} population formed in situ \cp{Ikoma2012}. It easier to form large planets further out, particularly beyond the snow-line where the increase in the local solid mass makes it easier to trigger runaway accretion to make a gas-giant. The relative scarcity of hot Jupiters found by \ct{Fressin2013} and \ct{Petigura2013}, is an indication that whatever migration mechanism brings in gas giants to orbits $\lesssim$100 days must be fairly rare.

One key puzzle, however, is the location of the break in the planet occurrence rate distribution. If it were due to a transition from a large rocky population to a sub-Neptune population, with planet occurrence declining with increase envelope fraction, then one would expect the break to occur at $\sim$1.5-1.8 $R_{\mathrm{\oplus}}$ the likely maximum size for bare rocky planets. Instead the break occurs at 2.8 $R_{\mathrm{\oplus}}$, indicating that the occurence plateau must include many volatile rich planets. Although 2.8 $R_{\mathrm{\oplus}}$ is far too large for bare rocky planets, it is achievable for H/He free water-worlds. A 10 $M_{\mathrm{\oplus}}$ planet with 80\% of its mass in a water envelope would be $\sim$2.7 $R_{\mathrm{\oplus}}$. As a result, it is at least possible that the break in the planet occurence distribution is a transition from an abundant population of rocky {\it and} water rich planets to a population with accreted H/He envelopes. Otherwise, models must explain why plateau should include a substantial population of planets with $\sim$1-3\% of their mass in H/He envelopes before dropping off at larger envelope fractions.

One potential explanation is that perhaps the $\sim$2-3 $R_{\mathrm{\oplus}}$ planets have hydrogen envelopes that were outgassed instead of accreted directly from the nebula. \ct{Elkins-Tanton2008} showed that low-mass planets can outgass up to $\sim$5\% of their mass after formation in H$_2$. However, this was only the case if the planets interiors we initially very wet, with $\sim$ half the mass of their initial mantles in water. This again requires a large of amount of water or other volatile ices to migrate to short period orbits.

It is also important to note that although the observed radius distribution may tell us the composition distribution of {\it Kepler} candidates today, this is not the same as the initial distribution the planets formed with. As showed in \ct{Lopez2012}, \ct{Lopez2013}, and \ct{Owen2013} the observed {\it Kepler} population has likely been significantly sculpted by photo-evaporation. Close-in low-mass planets have likely lost a significant fraction of their initial H/He inventories, resulting in smaller radii today. This effect is compounded by the fact that less irradiated planets should be able to accrete larger initial H/He envelopes in the first place \cp{Ikoma2012}. As more quarters of data are analyzed and the occurrence distribution pushes out to longer periods there should be a distinct increase in the abundance of Neptune and sub-Neptune sized planets.

Another potential effect of photo-evaporation is the opening up a slight ``occurrence valley'' in the radius-flux distribution \cp{Lopez2013,Owen2013}. Photo-evaporation makes it less likely that planets will survive with envelopes $\lesssim$1\% of their mass on highly irradiated orbits. Planets will tend to either retain more substantial envelope, or lose them entirely. More work needs to be done to carefully search for such a deficit, however there are some preliminary indications that it may exist. Examining both the raw candidate distribution \cp{Owen2013}, and a well-studied sample of M-dwarfs \cp{Morton2013}, appears to show a slight dip in the frequency of planets at $\sim$2 $R_{\mathrm{\oplus}}$. Such hints are still preliminary, but if real this has important implications for constraining the compositions of the {\it Kepler} population, since any large variation in the water fraction of close-in will tend to erase such a feature \cp{Lopez2013}. Using the models presented here, it is possible to instead study the {\it Kepler} envelope fraction distribution, which should aid in detecting any such ``occurrence valley.''

\section{Summary}

One of the key strengths of the thermal evolution models used here is that they allow us to predict the radius of a planet as a function of mostly observable parameters; namely, planet mass, incident flux, age, and composition. For Neptune and sub-Neptune size planets, we showed in section \ref{studysec}, the effect of varying planet mass or incident flux on the radius is an order of magnitude smaller than the effect of varying the fraction of a planet's mass in a H/He envelope. In section \ref{flatsec}, we described how this flatness in iso-composition mass-radius curves, arises as a natural result of our thermal evolution models. As a result of these features, planetary radius is to first order a proxy for the H/He inventory of sub-Neptune and larger planets, almost independent of their mass. In section \ref{compsec} we showed this close connection between radius and envelope fraction for the observed population of transiting planets with measured masses. We then demonstrated how our models allow us to recast the observed mass-radius distribution as a mass-{\it composition} relationship, allowing a more direct comparison to models of planet formation and evolution.

\acknowledgements{EDL would like to thank Angie Wolfgang, Jack Lissauer, Lauren Weiss, and Leslie Rogers for many helpful conversations. This research has made use of the Exoplanet Orbit Database and the Exoplanet Data Explorer at exoplanets.org." We acknowledge the support of NASA grant NNX09AC22G, NSF grant AST-1010017, and the UCSC Chancellor's Dissertion Year Fellowship.}

%\bibliographystyle{apj}
%\bibliography{myreferences}
\LongTables

\pagebreak

\begin{deluxetable*}{cccccccccc}[h!]
  \footnotesize
  \tablecaption{}
  \tablewidth{0pt}
  \tablehead{
  \colhead{Planet Name} & \colhead{Mass ($M_{\mathrm{\oplus}}$)} & \colhead{Radius ($R_{\mathrm{\oplus}}$)} & \colhead{H/He Envelope Fraction} 
}
  \startdata

Kepler-11b & 1.90$\pm^{1.40}_{1.00}$ & 1.80$\pm^{0.03}_{0.05}$ & 0.51\% $\pm^{0.46\%}_{0.36\%}$
\\\\
Kepler-11f & 2.00$\pm^{0.80}_{0.90}$ & 2.49$\pm^{0.04}_{0.07}$ & 4.0\% $\pm^{1.0\%}_{0.7\%}$
\\\\
Kepler-11c & 2.90$\pm^{2.90}_{1.60}$ & 2.87$\pm^{0.05}_{0.06}$ & 5.0\% $\pm^{1.1\%}_{0.8\%}$
\\\\
Kepler-36b & 4.46$\pm^{0.30}_{0.30}$ & 1.48$\pm^{0.03}_{0.03}$ & $>0.04\%$
\\\\
Kepler-10b & 4.51$\pm^{1.24}_{1.24}$ & 1.41$\pm^{0.03}_{0.03}$ & $>0.01\%$
\\\\
GJ1214b & 6.46$\pm^{0.99}_{0.99}$ & 2.67$\pm^{0.12}_{0.12}$ & 3.8\% $\pm^{1.3\%}_{1.1\%}$
\\\\
Kepler-18b & 6.87$\pm^{3.48}_{3.48}$ & 2.00$\pm^{0.09}_{0.09}$ & 0.31\% $\pm^{0.76\%}_{0.31\%}$
\\\\
Kepler-11d & 7.30$\pm^{0.80}_{1.50}$ & 3.12$\pm^{0.06}_{0.07}$ & 6.6\%$\pm^{1.3\%}_{1.2\%}$
\\\\
CoRoT-7b & 7.42$\pm^{1.21}_{1.21}$ & 1.67$\pm^{0.09}_{0.09}$ & $>0.03\%$
\\\\
Kepler-68b & 7.59$\pm^{2.06}_{2.06}$ & 2.30$\pm^{0.05}_{0.08}$ & 0.35\% $\pm^{0.38\%}_{0.17\%}$
\\\\
HD97658b & 7.86$\pm^{0.73}_{0.73}$ & 2.34$\pm^{0.18}_{0.15}$ & 0.99\% $\pm^{1.01\%}_{0.74\%}$
\\\\
Kepler-11e & 8.00$\pm^{1.50}_{2.10}$ & 4.19$\pm^{0.07}_{0.09}$ & 15.7\% $\pm^{1.7\%}_{1.7\%}$
\\\\
Kepler-36c & 8.10$\pm^{0.53}_{0.53}$ & 3.67$\pm^{0.05}_{0.05}$ & 8.6\% $\pm^{1.3\%}_{1.3\%}$
\\\\
55Cnce & 8.32$\pm^{0.39}_{0.39}$ & 1.99$\pm^{0.08}_{0.08}$ & 0.14\% $\pm^{0.21\%}_{0.13\%}$
\\\\
Kepler-20b & 8.45$\pm^{2.12}_{2.12}$ & 1.90$\pm^{0.11}_{0.20}$ & $>0.28\%$
\\\\
GJ3470b & 13.9$\pm^{1.63}_{1.63}$ & 4.19$\pm^{0.59}_{0.59}$ & 12.8\% $\pm^{5.2\%}_{5.0\%}$
\\\\
Kepler-20c & 15.7$\pm^{3.31}_{3.31}$ & 3.06$\pm^{0.19}_{0.30}$ & 3.5\% $\pm^{1.5\%}_{1.9\%}$
\\\\
Kepler-18d & 16.3$\pm^{1.39}_{1.39}$ & 6.97$\pm^{0.32}_{0.32}$ & 37.5\% $\pm^{3.5\%}_{3.7\%}$
\\\\
Kepler-18c & 17.2$\pm^{1.90}_{1.90}$ & 5.48$\pm^{0.25}_{0.25}$ & 23.6\% $\pm^{2.7\%}_{3.1\%}$
\\\\
HAT-P-26b & 18.6$\pm^{2.28}_{2.28}$ & 6.33$\pm^{0.58}_{0.58}$ & 31.7\% $\pm^{6.2\%}_{6.0\%}$
\\\\
GJ436b & 23.0$\pm^{1.01}_{1.01}$ & 4.22$\pm^{0.09}_{0.10}$ & 12.0\% $\pm^{1.2\%}_{2.1\%}$
\\\\
Kepler-4b & 24.5$\pm^{4.07}_{4.07}$ & 4.00$\pm^{0.21}_{0.21}$ & 7.7\% $\pm^{1.6\%}_{2.2\%}$
\\\\
HAT-P-11b & 26.2$\pm^{2.86}_{2.86}$ & 4.73$\pm^{0.15}_{0.15}$ & 15.1\% $\pm^{1.7\%}_{2.6\%}$
\\\\
Kepler-35b & 40.3$\pm^{6.35}_{6.35}$ & 8.16$\pm^{0.15}_{0.15}$ & 47.0\% $\pm^{2.0\%}_{3.0\%}$
\\\\
Kepler-9c & 53.5$\pm^{5.52}_{5.52}$ & 9.22$\pm^{0.75}_{0.75}$ & 64.9\% $\pm^{9.1\%}_{9.1\%}$
\\\\
HAT-P-18b & 62.6$\pm^{4.25}_{4.25}$ & 11.1$\pm^{0.58}_{0.58}$ & 87.1\% $\pm^{13.3\%}_{7.1\%}$
\\\\
HAT-P-12b & 66.9$\pm^{4.19}_{4.19}$ & 10.7$\pm^{0.32}_{0.23}$ & 80.3\% $\pm^{4.0\%}_{3.5\%}$
\\\\
CoRoT-8b & 68.6$\pm^{10.8}_{10.8}$ & 6.38$\pm^{0.22}_{0.22}$ & 28.6\% $\pm^{2.6\%}_{6.2\%}$
\\\\
Kepler-34b & 69.9$\pm^{3.49}_{3.17}$ & 8.56$\pm^{0.15}_{0.13}$ & 53.2\% $\pm^{1.9\%}_{4.1\%}$
\\\\
WASP-29b & 77.2$\pm^{6.39}_{6.39}$ & 8.87$\pm^{0.62}_{0.39}$ & 60.0\% $\pm^{7.6\%}_{6.1\%}$
\\\\
Kepler-9b & 79.0$\pm^{6.67}_{6.67}$ & 9.43$\pm^{0.77}_{0.77}$ & 62.5\% $\pm^{8.6\%}_{9.5\%}$
\\\\

  \enddata 
  \label{planettab}
  \tablecomments{Confirmed planets with well determined masses less than 100 $M_{\mathrm{\oplus}}$. Here we list each planets name, mass, radius, and the fraction of its mass in a H/He envelope according to our thermal evolution models. Planets with upper limits correspond to potentially rocky planets. The upper limit comes from the observed uncertainties on mass and radius and assuming a maximally iron rich core \ct{Marcus2010}}
\end{deluxetable*}

\pagebreak

\begin{deluxetable*}{ccccccccccccc}[h!]
  \footnotesize
  \tablecaption{Low Mass Planet Radii at 100 Myr, Solar Metallicity}
  \tablewidth{0pt}
  \tablehead{
  \colhead{Flux ($F_{\mathrm{\oplus}}$)} & \colhead{Mass ($M_{\mathrm{\oplus}}$)} & \colhead{0.01\%} & \colhead{0.02\%} & \colhead{0.05\%} & \colhead{0.1\%} & \colhead{0.2\%} & \colhead{0.5\%} & \colhead{1\%} & \colhead{2\%} & \colhead{5\%} & \colhead{10\%} & \colhead{20\%}
}
  \startdata
0.1 & 1 & 1.22 & 1.16 & 1.18 & 1.21 & 1.32 & 1.65 & 2.17 & 2.75 & 4.32 & 6.81 & 11.7
\\\\
0.1 & 1.5 & 1.30 & 1.24 & 1.26 & 1.30 & 1.40 & 1.71 & 2.15 & 2.65 & 3.97 & 6.18 & 10.6
\\\\
0.1 & 2.4 & 1.41 & 1.36 & 1.40 & 1.42 & 1.53 & 1.79 & 2.17 & 2.58 & 3.66 & 5.36 & 9.05
\\\\
0.1 & 3.6 & 1.53 & 1.49 & 1.51 & 1.54 & 1.64 & 1.89 & 2.21 & 2.56 & 3.49 & 4.93 & 7.86
\\\\
0.1 & 5.5 & 1.66 & 1.63 & 1.66 & 1.69 & 1.79 & 2.01 & 2.28 & 2.60 & 3.37 & 4.58 & 6.96
\\\\
0.1 & 8.5 & 1.81 & 1.79 & 1.82 & 1.85 & 1.95 & 2.14 & 2.39 & 2.67 & 3.36 & 4.35 & 6.32
\\\\
0.1 & 13 & 1.97 & 1.97 & 1.98 & 2.02 & 2.11 & 2.30 & 2.52 & 2.78 & 3.41 & 4.29 & 5.94
\\\\
0.1 & 20 & 2.15 & 2.15 & 2.17 & 2.20 & 2.29 & 2.47 & 2.67 & 2.93 & 3.52 & 4.32 & 5.75
\\\\
10 & 1 & 1.32 & 1.24 & 1.27 & 1.31 & 1.44 & 1.82 & 2.40 & 3.06 & 4.72 & 7.13 & 11.1
\\\\
10 & 1.5 & 1.36 & 1.32 & 1.35 & 1.38 & 1.50 & 1.84 & 2.32 & 2.88 & 4.31 & 6.47 & 10.4
\\\\
10 & 2.4 & 1.46 & 1.43 & 1.48 & 1.50 & 1.59 & 1.88 & 2.26 & 2.71 & 3.88 & 5.67 & 9.14
\\\\
10 & 3.6 & 1.57 & 1.55 & 1.58 & 1.60 & 1.71 & 1.95 & 2.27 & 2.64 & 3.61 & 5.13 & 8.11
\\\\
10 & 5.5 & 1.69 & 1.68 & 1.71 & 1.73 & 1.84 & 2.05 & 2.33 & 2.66 & 3.46 & 4.70 & 7.13
\\\\
10 & 8.5 & 1.84 & 1.83 & 1.86 & 1.89 & 1.98 & 2.18 & 2.43 & 2.72 & 3.42 & 4.43 & 6.39
\\\\
10 & 13 & 1.99 & 2.01 & 2.02 & 2.05 & 2.14 & 2.32 & 2.55 & 2.82 & 3.46 & 4.35 & 5.96
\\\\
10 & 20 & 2.17 & 2.18 & 2.19 & 2.23 & 2.31 & 2.49 & 2.69 & 2.95 & 3.56 & 4.37 & 5.77
\\\\
1000 & 1 & 1.59 & 1.63 & 1.70 & 1.75 & 1.83 & 2.30 & 3.12 & 3.99 & 6.21 & 8.88 & 11.3
\\\\
1000 & 1.5 & 1.63 & 1.67 & 1.72 & 1.77 & 1.89 & 2.31 & 3.02 & 3.83 & 6.01 & 9.41 & 14.0
\\\\
1000 & 2.4 & 1.70 & 1.72 & 1.77 & 1.81 & 1.93 & 2.32 & 2.90 & 3.55 & 5.35 & 8.59 & 15.4
\\\\
1000 & 3.6 & 1.77 & 1.79 & 1.83 & 1.87 & 1.99 & 2.34 & 2.81 & 3.36 & 4.82 & 7.27 & 13.4
\\\\
1000 & 5.5 & 1.87 & 1.88 & 1.92 & 1.96 & 2.08 & 2.37 & 2.76 & 3.22 & 4.39 & 6.25 & 10.3
\\\\
1000 & 8.5 & 1.99 & 2.00 & 2.03 & 2.08 & 2.19 & 2.50 & 2.76 & 3.15 & 4.12 & 5.56 & 8.48
\\\\
1000 & 13 & 2.12 & 2.12 & 2.15 & 2.21 & 2.31 & 2.58 & 2.81 & 3.16 & 3.99 & 5.18 & 7.43
\\\\
1000 & 20 & 2.27 & 2.27 & 2.30 & 2.35 & 2.45 & 2.68 & 2.90 & 3.21 & 3.94 & 4.97 & 6.80
\\\\
  \enddata 
  \label{sol100tab}
  \tablecomments{Radii of planets, in $R_{\mathrm{\oplus}}$. Column 1 is incident flux on the planet, relative to the solar constant. Column 2 is the total planet mass in $M_{\mathrm{\oplus}}$. Otherwise, column headers indicate the fraction of a planet's mass in the H/He envelope.}
\end{deluxetable*}
\pagebreak
\begin{deluxetable*}{ccccccccccccc}[h!]
  \footnotesize
  \tablecaption{Low Mass Planet Radii at 1 Gyr, Solar Metallicity}
  \tablewidth{0pt}
  \tablehead{
  \colhead{Flux ($F_{\mathrm{\oplus}}$)} & \colhead{Mass ($M_{\mathrm{\oplus}}$)} & \colhead{0.01\%} & \colhead{0.02\%} & \colhead{0.05\%} & \colhead{0.1\%} & \colhead{0.2\%} & \colhead{0.5\%} & \colhead{1\%} & \colhead{2\%} & \colhead{5\%} & \colhead{10\%} & \colhead{20\%}
}
  \startdata
0.1 & 1 & 1.07 & 1.09 & 1.12 & 1.15 & 1.28 & 1.55 & 1.79 & 2.13 & 2.98 & 4.26 & 6.74
\\\\
0.1 & 1.5 & 1.18 & 1.19 & 1.22 & 1.26 & 1.38 & 1.62 & 1.82 & 2.13 & 2.87 & 3.96 & 6.10
\\\\
0.1 & 2.4 & 1.32 & 1.33 & 1.36 & 1.39 & 1.52 & 1.72 & 1.90 & 2.16 & 2.81 & 3.75 & 5.52
\\\\
0.1 & 3.6 & 1.45 & 1.46 & 1.49 & 1.52 & 1.65 & 1.82 & 1.99 & 2.23 & 2.81 & 3.65 & 5.21
\\\\
0.1 & 5.5 & 1.60 & 1.61 & 1.64 & 1.67 & 1.79 & 1.95 & 2.11 & 2.34 & 2.87 & 3.62 & 5.00
\\\\
0.1 & 8.5 & 1.77 & 1.78 & 1.80 & 1.83 & 1.94 & 2.10 & 2.25 & 2.47 & 2.97 & 3.67 & 4.91
\\\\
0.1 & 13 & 1.94 & 1.95 & 1.97 & 2.00 & 2.11 & 2.25 & 2.40 & 2.61 & 3.11 & 3.77 & 4.92
\\\\
0.1 & 20 & 2.12 & 2.13 & 2.16 & 2.19 & 2.30 & 2.42 & 2.57 & 2.78 & 3.28 & 3.92 & 5.00
\\\\
10 & 1 & 1.18 & 1.20 & 1.23 & 1.27 & 1.47 & 1.81 & 2.12 & 2.58 & 3.63 & 5.07 & 7.45
\\\\
10 & 1.5 & 1.27 & 1.29 & 1.32 & 1.36 & 1.52 & 1.82 & 2.08 & 2.47 & 3.40 & 4.68 & 6.96
\\\\
10 & 2.4 & 1.40 & 1.41 & 1.44 & 1.48 & 1.63 & 1.86 & 2.08 & 2.41 & 3.18 & 4.26 & 6.24
\\\\
10 & 3.6 & 1.51 & 1.53 & 1.55 & 1.59 & 1.72 & 1.93 & 2.12 & 2.40 & 3.07 & 4.02 & 5.73
\\\\
10 & 5.5 & 1.65 & 1.66 & 1.69 & 1.72 & 1.85 & 2.02 & 2.19 & 2.45 & 3.04 & 3.86 & 5.34
\\\\
10 & 8.5 & 1.81 & 1.82 & 1.84 & 1.88 & 1.99 & 2.15 & 2.31 & 2.54 & 3.08 & 3.81 & 5.09
\\\\
10 & 13 & 1.97 & 1.98 & 2.01 & 2.04 & 2.15 & 2.29 & 2.44 & 2.67 & 3.18 & 3.86 & 5.02
\\\\
10 & 20 & 2.15 & 2.16 & 2.18 & 2.22 & 2.32 & 2.45 & 2.60 & 2.82 & 3.33 & 3.99 & 5.07
\\\\
1000 & 1 & 1.61 & 1.65 & 1.71 & 1.77 & 1.81 & 2.15 & 2.50 & 3.01 & 4.24 & 6.04 & 8.75
\\\\
1000 & 1.5 & 1.65 & 1.68 & 1.73 & 1.78 & 1.87 & 2.18 & 2.50 & 2.98 & 4.14 & 5.91 & 9.34
\\\\
1000 & 2.4 & 1.71 & 1.73 & 1.78 & 1.82 & 1.93 & 2.21 & 2.50 & 2.91 & 3.93 & 5.50 & 8.76
\\\\
1000 & 3.6 & 1.78 & 1.80 & 1.84 & 1.87 & 1.99 & 2.24 & 2.50 & 2.87 & 3.77 & 5.11 & 7.86
\\\\
1000 & 5.5 & 1.87 & 1.89 & 1.92 & 1.94 & 2.10 & 2.30 & 2.52 & 2.85 & 3.65 & 4.79 & 7.00
\\\\
1000 & 8.5 & 1.99 & 2.00 & 2.02 & 2.05 & 2.19 & 2.38 & 2.58 & 2.88 & 3.59 & 4.58 & 6.39
\\\\
1000 & 13 & 2.12 & 2.13 & 2.15 & 2.19 & 2.31 & 2.48 & 2.66 & 2.94 & 3.59 & 4.48 & 6.05
\\\\
1000 & 20 & 2.27 & 2.27 & 2.29 & 2.34 & 2.45 & 2.61 & 2.78 & 3.04 & 3.65 & 4.47 & 5.85
\\\\
  \enddata 
  \label{sol1tab}
  \tablecomments{Radii of planets, in $R_{\mathrm{\oplus}}$. Column 1 is incident flux on the planet, relative to the solar constant. Column 2 is the total planet mass in $M_{\mathrm{\oplus}}$. Otherwise, column headers indicate the fraction of a planet's mass in the H/He envelope.}
\end{deluxetable*}
\pagebreak
\begin{deluxetable*}{ccccccccccccc}[h!]
  \footnotesize
  \tablecaption{Low Mass Planet Radii at 10 Gyr, Solar Metallicity}
  \tablewidth{0pt}
  \tablehead{
  \colhead{Flux ($F_{\mathrm{\oplus}}$)} & \colhead{Mass ($M_{\mathrm{\oplus}}$)} & \colhead{0.01\%} & \colhead{0.02\%} & \colhead{0.05\%} & \colhead{0.1\%} & \colhead{0.2\%} & \colhead{0.5\%} & \colhead{1\%} & \colhead{2\%} & \colhead{5\%} & \colhead{10\%} & \colhead{20\%}
}
  \startdata
0.1 & 1 & 1.08 & 1.10 & 1.13 & 1.17 & 1.22 & 1.37 & 1.53 & 1.75 & 2.25 & 2.94 & 4.14
\\\\
0.1 & 1.5 & 1.19 & 1.20 & 1.23 & 1.27 & 1.31 & 1.45 & 1.60 & 1.81 & 2.28 & 2.93 & 4.05
\\\\
0.1 & 2.4 & 1.32 & 1.34 & 1.37 & 1.40 & 1.45 & 1.58 & 1.71 & 1.90 & 2.35 & 2.95 & 3.98
\\\\
0.1 & 3.6 & 1.45 & 1.47 & 1.49 & 1.53 & 1.58 & 1.70 & 1.82 & 2.01 & 2.44 & 3.01 & 3.97
\\\\
0.1 & 5.5 & 1.60 & 1.62 & 1.64 & 1.67 & 1.75 & 1.84 & 1.96 & 2.15 & 2.56 & 3.11 & 4.03
\\\\
0.1 & 8.5 & 1.77 & 1.78 & 1.80 & 1.84 & 1.91 & 2.00 & 2.13 & 2.31 & 2.72 & 3.25 & 4.14
\\\\
0.1 & 13 & 1.94 & 1.95 & 1.97 & 2.00 & 2.09 & 2.17 & 2.30 & 2.48 & 2.90 & 3.44 & 4.31
\\\\
0.1 & 20 & 2.12 & 2.14 & 2.16 & 2.19 & 2.25 & 2.36 & 2.49 & 2.68 & 3.10 & 3.65 & 4.53
\\\\
10 & 1 & 1.23 & 1.25 & 1.28 & 1.31 & 1.44 & 1.68 & 1.87 & 2.17 & 2.84 & 3.70 & 5.11
\\\\
10 & 1.5 & 1.31 & 1.33 & 1.36 & 1.40 & 1.49 & 1.72 & 1.90 & 2.19 & 2.83 & 3.66 & 5.03
\\\\
10 & 2.4 & 1.43 & 1.44 & 1.47 & 1.51 & 1.60 & 1.78 & 1.96 & 2.21 & 2.80 & 3.58 & 4.89
\\\\
10 & 3.6 & 1.54 & 1.55 & 1.58 & 1.62 & 1.73 & 1.87 & 2.03 & 2.27 & 2.81 & 3.53 & 4.75
\\\\
10 & 5.5 & 1.67 & 1.69 & 1.71 & 1.75 & 1.85 & 1.98 & 2.13 & 2.35 & 2.86 & 3.52 & 4.64
\\\\
10 & 8.5 & 1.82 & 1.84 & 1.86 & 1.90 & 1.98 & 2.11 & 2.25 & 2.47 & 2.95 & 3.58 & 4.61
\\\\
10 & 13 & 1.98 & 1.99 & 2.02 & 2.05 & 2.13 & 2.26 & 2.40 & 2.61 & 3.07 & 3.68 & 4.66
\\\\
10 & 20 & 2.16 & 2.17 & 2.20 & 2.23 & 2.32 & 2.43 & 2.56 & 2.77 & 3.23 & 3.83 & 4.77
\\\\
1000 & 1 & 1.76 & 1.81 & 1.88 & 1.96 & 2.01 & 2.08 & 2.18 & 2.31 & 2.70 & 3.49 & 4.88
\\\\
1000 & 1.5 & 1.77 & 1.81 & 1.88 & 1.94 & 1.99 & 2.08 & 2.17 & 2.33 & 2.91 & 3.76 & 5.36
\\\\
1000 & 2.4 & 1.82 & 1.85 & 1.90 & 1.95 & 2.00 & 2.08 & 2.22 & 2.49 & 3.10 & 3.94 & 5.55
\\\\
1000 & 3.6 & 1.87 & 1.90 & 1.94 & 1.98 & 2.03 & 2.12 & 2.30 & 2.58 & 3.20 & 4.03 & 5.54
\\\\
1000 & 5.5 & 1.95 & 1.97 & 2.01 & 2.04 & 2.10 & 2.21 & 2.38 & 2.64 & 3.26 & 4.08 & 5.49
\\\\
1000 & 8.5 & 2.05 & 2.07 & 2.10 & 2.12 & 2.19 & 2.31 & 2.48 & 2.73 & 3.31 & 4.10 & 5.44
\\\\
1000 & 13 & 2.17 & 2.18 & 2.21 & 2.23 & 2.34 & 2.43 & 2.59 & 2.83 & 3.38 & 4.13 & 5.40
\\\\
1000 & 20 & 2.31 & 2.32 & 2.34 & 2.36 & 2.47 & 2.57 & 2.72 & 2.95 & 3.49 & 4.20 & 5.38
\\\\

  \enddata 
  \label{sol10tab}
  \tablecomments{Radii of planets, in $R_{\mathrm{\oplus}}$. Column 1 is incident flux on the planet, relative to the solar constant. Column 2 is the total planet mass in $M_{\mathrm{\oplus}}$. Otherwise, column headers indicate the fraction of a planet's mass in the H/He envelope.}
\end{deluxetable*}
\pagebreak
\begin{deluxetable*}{ccccccccccccc}[h!]
  \footnotesize
  \tablecaption{Low Mass Planet Radii at 100 Myr, Enhanced Opacity}
  \tablewidth{0pt}
  \tablehead{
  \colhead{Flux ($F_{\mathrm{\oplus}}$)} & \colhead{Mass ($M_{\mathrm{\oplus}}$)} & \colhead{0.01\%} & \colhead{0.02\%} & \colhead{0.05\%} & \colhead{0.1\%} & \colhead{0.2\%} & \colhead{0.5\%} & \colhead{1\%} & \colhead{2\%} & \colhead{5\%} & \colhead{10\%} & \colhead{20\%}
}
  \startdata
0.1 & 1 & 1.25 & 1.17 & 1.19 & 1.24 & 1.43 & 1.89 & 2.57 & 3.00 & 4.32 & 6.81 & 11.7
\\\\
0.1 & 1.5 & 1.31 & 1.25 & 1.28 & 1.31 & 1.47 & 1.93 & 2.58 & 3.17 & 3.97 & 6.18 & 10.6
\\\\
0.1 & 2.4 & 1.42 & 1.37 & 1.41 & 1.44 & 1.60 & 2.14 & 2.51 & 3.11 & 4.30 & 5.36 & 9.05
\\\\
0.1 & 3.6 & 1.53 & 1.49 & 1.52 & 1.57 & 1.71 & 2.18 & 2.51 & 3.02 & 4.31 & 5.57 & 7.86
\\\\
0.1 & 5.5 & 1.66 & 1.63 & 1.67 & 1.72 & 1.84 & 2.23 & 2.53 & 2.97 & 4.09 & 5.69 & 7.25
\\\\
0.1 & 8.5 & 1.82 & 1.79 & 1.82 & 1.87 & 2.00 & 2.33 & 2.57 & 2.95 & 3.90 & 5.29 & 7.73
\\\\
0.1 & 13 & 1.97 & 1.96 & 1.99 & 2.04 & 2.16 & 2.45 & 2.67 & 3.00 & 3.81 & 4.99 & 7.25
\\\\
0.1 & 20 & 2.15 & 2.14 & 2.17 & 2.23 & 2.33 & 2.58 & 2.79 & 3.10 & 3.82 & 4.82 & 6.68
\\\\
10 & 1 & 1.34 & 1.25 & 1.29 & 1.35 & 1.53 & 2.05 & 2.79 & 3.12 & 4.72 & 7.13 & 11.1
\\\\
10 & 1.5 & 1.38 & 1.33 & 1.37 & 1.41 & 1.58 & 2.07 & 2.80 & 3.41 & 4.31 & 6.47 & 10.4
\\\\
10 & 2.4 & 1.47 & 1.44 & 1.48 & 1.51 & 1.68 & 2.25 & 2.67 & 3.32 & 4.50 & 5.67 & 9.14
\\\\
10 & 3.6 & 1.58 & 1.54 & 1.58 & 1.64 & 1.77 & 2.26 & 2.62 & 3.17 & 4.55 & 5.71 & 8.11
\\\\
10 & 5.5 & 1.70 & 1.67 & 1.71 & 1.77 & 1.89 & 2.30 & 2.60 & 3.07 & 4.27 & 5.94 & 7.30
\\\\
10 & 8.5 & 1.85 & 1.83 & 1.86 & 1.92 & 2.04 & 2.37 & 2.63 & 3.01 & 4.01 & 5.50 & 7.98
\\\\
10 & 13 & 1.99 & 1.99 & 2.02 & 2.08 & 2.19 & 2.48 & 2.70 & 3.04 & 3.88 & 5.12 & 7.50
\\\\
10 & 20 & 2.17 & 2.17 & 2.19 & 2.25 & 2.35 & 2.61 & 2.82 & 3.13 & 3.87 & 4.90 & 6.81
\\\\
1000 & 1 & 1.59 & 1.63 & 1.70 & 1.75 & 1.88 & 2.42 & 3.13 & 3.99 & 6.21 & 8.88 & 11.3
\\\\
1000 & 1.5 & 1.63 & 1.67 & 1.72 & 1.77 & 1.90 & 2.46 & 3.25 & 3.84 & 6.01 & 9.41 & 14.0
\\\\
1000 & 2.4 & 1.70 & 1.72 & 1.77 & 1.81 & 1.97 & 2.63 & 3.13 & 3.89 & 5.35 & 8.59 & 15.4
\\\\
1000 & 3.6 & 1.77 & 1.79 & 1.83 & 1.87 & 2.02 & 2.57 & 3.01 & 3.67 & 5.23 & 7.27 & 13.4
\\\\
1000 & 5.5 & 1.87 & 1.88 & 1.92 & 1.96 & 2.10 & 2.54 & 2.90 & 3.46 & 4.89 & 6.68 & 10.3
\\\\
1000 & 8.5 & 1.99 & 2.00 & 2.03 & 2.08 & 2.21 & 2.56 & 2.86 & 3.31 & 4.48 & 6.23 & 8.82
\\\\
1000 & 13 & 2.12 & 2.12 & 2.15 & 2.21 & 2.32 & 2.63 & 2.89 & 3.27 & 4.22 & 5.64 & 8.36
\\\\
1000 & 20 & 2.27 & 2.27 & 2.30 & 2.36 & 2.47 & 2.73 & 2.95 & 3.30 & 4.12 & 5.28 & 7.38
\\\\
  \enddata 
  \label{sol100tab50}
  \tablecomments{Radii of planets, in $R_{\mathrm{\oplus}}$. Column 1 is incident flux on the planet, relative to the solar constant. Column 2 is the total planet mass in $M_{\mathrm{\oplus}}$. Otherwise, column headers indicate the fraction of a planet's mass in the H/He envelope.}
\end{deluxetable*}
\pagebreak
\begin{deluxetable*}{ccccccccccccc}[h!]
  \footnotesize
  \tablecaption{Low Mass Planet Radii at 1 Gyr, Enhanced Opacity}
  \tablewidth{0pt}
  \tablehead{
  \colhead{Flux ($F_{\mathrm{\oplus}}$)} & \colhead{Mass ($M_{\mathrm{\oplus}}$)} & \colhead{0.01\%} & \colhead{0.02\%} & \colhead{0.05\%} & \colhead{0.1\%} & \colhead{0.2\%} & \colhead{0.5\%} & \colhead{1\%} & \colhead{2\%} & \colhead{5\%} & \colhead{10\%} & \colhead{20\%} 
}
  \startdata
0.1 & 1 & 1.07 & 1.09 & 1.12 & 1.15 & 1.37 & 1.68 & 1.98 & 2.43 & 3.11 & 4.26 & 6.74
\\\\
0.1 & 1.5 & 1.18 & 1.19 & 1.22 & 1.26 & 1.45 & 1.74 & 1.99 & 2.38 & 3.27 & 3.96 & 6.10
\\\\
0.1 & 2.4 & 1.32 & 1.33 & 1.36 & 1.39 & 1.60 & 1.82 & 2.05 & 2.38 & 3.22 & 4.23 & 5.52
\\\\
0.1 & 3.6 & 1.45 & 1.46 & 1.49 & 1.52 & 1.71 & 1.90 & 2.10 & 2.42 & 3.15 & 4.25 & 5.62
\\\\
0.1 & 5.5 & 1.60 & 1.61 & 1.64 & 1.67 & 1.85 & 2.02 & 2.19 & 2.47 & 3.16 & 4.13 & 5.84
\\\\
0.1 & 8.5 & 1.77 & 1.78 & 1.80 & 1.84 & 1.99 & 2.15 & 2.32 & 2.57 & 3.19 & 4.09 & 5.74
\\\\
0.1 & 13 & 1.94 & 1.95 & 1.97 & 2.01 & 2.15 & 2.30 & 2.46 & 2.70 & 3.28 & 4.11 & 5.58
\\\\
0.1 & 20 & 2.12 & 2.13 & 2.16 & 2.21 & 2.32 & 2.46 & 2.62 & 2.86 & 3.42 & 4.17 & 5.56
\\\\
10 & 1 & 1.18 & 1.20 & 1.23 & 1.27 & 1.50 & 1.84 & 2.20 & 2.72 & 3.63 & 5.07 & 7.45
\\\\
10 & 1.5 & 1.27 & 1.29 & 1.32 & 1.36 & 1.56 & 1.87 & 2.16 & 2.60 & 3.58 & 4.68 & 6.96
\\\\
10 & 2.4 & 1.40 & 1.41 & 1.44 & 1.48 & 1.68 & 1.92 & 2.17 & 2.54 & 3.47 & 4.52 & 6.24
\\\\
10 & 3.6 & 1.51 & 1.53 & 1.55 & 1.59 & 1.77 & 1.98 & 2.20 & 2.54 & 3.34 & 4.53 & 5.85
\\\\
10 & 5.5 & 1.65 & 1.66 & 1.69 & 1.72 & 1.90 & 2.07 & 2.27 & 2.56 & 3.31 & 4.36 & 6.14
\\\\
10 & 8.5 & 1.81 & 1.82 & 1.84 & 1.89 & 2.03 & 2.20 & 2.37 & 2.64 & 3.31 & 4.27 & 5.99
\\\\
10 & 13 & 1.97 & 1.98 & 2.01 & 2.05 & 2.18 & 2.34 & 2.50 & 2.76 & 3.37 & 4.24 & 5.76
\\\\
10 & 20 & 2.15 & 2.16 & 2.18 & 2.24 & 2.36 & 2.49 & 2.65 & 2.90 & 3.49 & 4.27 & 5.68
\\\\
1000 & 1 & 1.61 & 1.65 & 1.71 & 1.77 & 1.85 & 2.20 & 2.60 & 3.09 & 4.24 & 6.04 & 8.75
\\\\
1000 & 1.5 & 1.65 & 1.68 & 1.73 & 1.78 & 1.88 & 2.24 & 2.59 & 3.10 & 4.14 & 5.91 & 9.34
\\\\
1000 & 2.4 & 1.71 & 1.73 & 1.78 & 1.82 & 1.96 & 2.26 & 2.57 & 3.02 & 4.13 & 5.50 & 8.76
\\\\
1000 & 3.6 & 1.78 & 1.80 & 1.84 & 1.87 & 2.02 & 2.27 & 2.54 & 2.96 & 3.94 & 5.34 & 7.86
\\\\
1000 & 5.5 & 1.87 & 1.89 & 1.92 & 1.94 & 2.11 & 2.32 & 2.54 & 2.90 & 3.80 & 5.05 & 7.13
\\\\
1000 & 8.5 & 1.99 & 2.00 & 2.02 & 2.05 & 2.21 & 2.39 & 2.59 & 2.90 & 3.68 & 4.81 & 6.84
\\\\
1000 & 13 & 2.12 & 2.13 & 2.15 & 2.19 & 2.32 & 2.49 & 2.67 & 2.96 & 3.65 & 4.65 & 6.41
\\\\
1000 & 20 & 2.27 & 2.27 & 2.29 & 2.34 & 2.46 & 2.61 & 2.78 & 3.06 & 3.70 & 4.57 & 6.14
\\\\
  \enddata 
  \label{sol1tab50}
  \tablecomments{Radii of planets, in $R_{\mathrm{\oplus}}$. Column 1 is incident flux on the planet, relative to the solar constant. Column 2 is the total planet mass in $M_{\mathrm{\oplus}}$. Otherwise, column headers indicate the fraction of a planet's mass in the H/He envelope.}
\end{deluxetable*}
\pagebreak
\begin{deluxetable*}{ccccccccccccccccc}[h!]
  \footnotesize
  \tablecaption{Low Mass Planet Radii at 10 Gyr, Enhanced Opacity}
  \tablewidth{0pt}
  \tablehead{
  \colhead{Flux ($F_{\mathrm{\oplus}}$)} & \colhead{Mass ($M_{\mathrm{\oplus}}$)} & \colhead{0.01\%} & \colhead{0.02\%} & \colhead{0.05\%} & \colhead{0.1\%} & \colhead{0.2\%} & \colhead{0.5\%} & \colhead{1\%} & \colhead{2\%} & \colhead{5\%} & \colhead{10\%} & \colhead{20\%} 
}
  \startdata
0.1 & 1 & 1.08 & 1.10 & 1.13 & 1.17 & 1.24 & 1.37 & 1.53 & 1.75 & 2.32 & 2.97 & 4.14
\\\\
0.1 & 1.5 & 1.19 & 1.20 & 1.23 & 1.27 & 1.34 & 1.46 & 1.61 & 1.82 & 2.33 & 3.04 & 4.05
\\\\
0.1 & 2.4 & 1.32 & 1.34 & 1.37 & 1.40 & 1.47 & 1.57 & 1.71 & 1.92 & 2.40 & 3.07 & 4.13
\\\\
0.1 & 3.6 & 1.45 & 1.47 & 1.49 & 1.53 & 1.59 & 1.70 & 1.83 & 2.01 & 2.49 & 3.12 & 4.23
\\\\
0.1 & 5.5 & 1.60 & 1.62 & 1.64 & 1.67 & 1.76 & 1.85 & 1.97 & 2.16 & 2.59 & 3.22 & 4.27
\\\\
0.1 & 8.5 & 1.77 & 1.78 & 1.80 & 1.84 & 1.88 & 2.02 & 2.14 & 2.32 & 2.75 & 3.34 & 4.37
\\\\
0.1 & 13 & 1.94 & 1.95 & 1.97 & 2.00 & 2.10 & 2.18 & 2.30 & 2.50 & 2.93 & 3.50 & 4.51
\\\\
0.1 & 20 & 2.12 & 2.14 & 2.16 & 2.19 & 2.27 & 2.37 & 2.49 & 2.69 & 3.13 & 3.71 & 4.66
\\\\
10 & 1 & 1.23 & 1.25 & 1.28 & 1.31 & 1.39 & 1.55 & 1.75 & 2.02 & 2.72 & 3.70 & 5.11
\\\\
10 & 1.5 & 1.31 & 1.33 & 1.36 & 1.40 & 1.48 & 1.62 & 1.79 & 2.05 & 2.66 & 3.57 & 5.03
\\\\
10 & 2.4 & 1.43 & 1.44 & 1.47 & 1.51 & 1.58 & 1.70 & 1.85 & 2.10 & 2.65 & 3.44 & 4.89
\\\\
10 & 3.6 & 1.54 & 1.55 & 1.58 & 1.62 & 1.69 & 1.80 & 1.94 & 2.15 & 2.69 & 3.41 & 4.67
\\\\
10 & 5.5 & 1.67 & 1.69 & 1.71 & 1.75 & 1.84 & 1.93 & 2.07 & 2.27 & 2.75 & 3.44 & 4.61
\\\\
10 & 8.5 & 1.82 & 1.84 & 1.86 & 1.90 & 1.98 & 2.08 & 2.21 & 2.41 & 2.86 & 3.51 & 4.62
\\\\
10 & 13 & 1.98 & 1.99 & 2.02 & 2.05 & 2.14 & 2.23 & 2.36 & 2.56 & 3.02 & 3.63 & 4.70
\\\\
10 & 20 & 2.16 & 2.17 & 2.20 & 2.23 & 2.31 & 2.41 & 2.54 & 2.74 & 3.21 & 3.81 & 4.81
\\\\
1000 & 1 & 1.76 & 1.81 & 1.88 & 1.96 & 2.01 & 2.08 & 2.16 & 2.29 & 2.60 & 3.49 & 4.88
\\\\
1000 & 1.5 & 1.77 & 1.81 & 1.88 & 1.94 & 1.99 & 2.05 & 2.15 & 2.28 & 2.83 & 3.72 & 5.36
\\\\
1000 & 2.4 & 1.82 & 1.85 & 1.90 & 1.95 & 1.99 & 2.05 & 2.14 & 2.36 & 3.01 & 3.89 & 5.55
\\\\
1000 & 3.6 & 1.87 & 1.90 & 1.94 & 1.98 & 2.02 & 2.09 & 2.18 & 2.42 & 3.10 & 3.97 & 5.48
\\\\
1000 & 5.5 & 1.95 & 1.97 & 2.01 & 2.04 & 2.09 & 2.15 & 2.31 & 2.54 & 3.14 & 4.01 & 5.47
\\\\
1000 & 8.5 & 2.05 & 2.07 & 2.10 & 2.12 & 2.15 & 2.28 & 2.43 & 2.67 & 3.21 & 4.01 & 5.40
\\\\
1000 & 13 & 2.17 & 2.18 & 2.21 & 2.23 & 2.28 & 2.40 & 2.55 & 2.79 & 3.33 & 4.05 & 5.35
\\\\
1000 & 20 & 2.31 & 2.32 & 2.34 & 2.36 & 2.45 & 2.55 & 2.69 & 2.92 & 3.45 & 4.15 & 5.32
\\\\

  \enddata 
  \label{sol10tab50}
  \tablecomments{Radii of planets, in $R_{\mathrm{\oplus}}$. Column 1 is incident flux on the planet, relative to the solar constant. Column 2 is the total planet mass in $M_{\mathrm{\oplus}}$. Otherwise, column headers indicate the fraction of a planet's mass in the H/He envelope.}
\end{deluxetable*}

\end{document}